\documentclass[11pt,prd,a4paper,preprintnumbers,amsmath,amssymb,nofootinbib]{article}
\usepackage{feynmp-auto,expdlist}
\usepackage{amsmath,amsfonts,amssymb}
\usepackage{graphicx}
\usepackage{enumerate}
\usepackage{hyperref}
\usepackage{latexsym}
\usepackage{hepnicenames}
\usepackage{enumerate}
\usepackage{soul}
\usepackage[normalem]{ulem}
\usepackage{wasysym}
\usepackage{makecell}
\usepackage{bbm}
\usepackage{physics}
\usepackage[version=4]{mhchem}
\usepackage{amsmath}

\font\tenrsfs=rsfs10 at 12pt
\font\sevenrsfs=rsfs7
\font\fiversfs=rsfs5
\newfam\rsfsfam
\textfont\rsfsfam=\tenrsfs
\scriptfont\rsfsfam=\sevenrsfs
\scriptscriptfont\rsfsfam=\fiversfs

\numberwithin{equation}{section}

\usepackage{mathrsfs}
\usepackage{braket}
\usepackage{titling}
\usepackage{amsmath}
\usepackage{slashed}
\usepackage{amssymb}
\usepackage{epsfig}
\usepackage{graphicx}
\usepackage{color}
\usepackage{rotating}
\usepackage{hyperref}
\usepackage[margin=1.in]{geometry}
\usepackage[table,xcdraw,dvipsnames]{xcolor}
\usepackage[compress,numbers,sort]{natbib}
\usepackage{colortbl}
\usepackage{pdflscape}
\usepackage{color}
\usepackage{mathtools}
\usepackage{colortbl}
\usepackage{comment}
\usepackage{multirow}
\usepackage{booktabs}

\newcommand{\A}{{\cal A}}
\newcommand{\B}{{\cal B}}

\newcommand{\X}{{\cal X}}

\newcommand{\SU}{{\rm SU}}

\newcommand{\U}{{\rm U}}

\newcommand{\lag}{{\mathscr L}}
\newcommand{\N}{{\cal N}}

\newcommand{\Y}{{\cal Y}}

\newcommand{\BR}{{\mathcal B}}

\definecolor{nicered}{rgb}{0.7,0.1,0.1}
\definecolor{nicegreen}{rgb}{0.1,0.5,0.1}
\definecolor{red}{rgb}{1.0, 0, 0}
\definecolor{niceblue}{rgb}{0,0,0.8}
\allowdisplaybreaks

\definecolor{blus}{cmyk}{1,1,0,0.6}
\definecolor{verde}{cmyk}{0.92,0,0.59,0.25}
\definecolor{rossos}{cmyk}{0,1,1,0.55}
\definecolor{piggypink}{rgb}{0.99, 0.87, 0.9}

\hypersetup{colorlinks,bookmarksopen,bookmarksnumbered,linkcolor=blus,pdfstartview=FitH,urlcolor=rossos,citecolor=verde}

\def\eq#1{{Eq.~(\ref{#1})}}
\def\eqs#1#2{{Eqs.~(\ref{#1})--(\ref{#2})}}
\def\fig#1{{Fig.~\ref{#1}}}

\def\Table#1{{Table~\ref{#1}}}

\def\sect#1{{Section~\ref{#1}}}

\def\vev#1{\left\langle #1\right\rangle}

\renewcommand{\bar}{\overline}

\newcommand{\beq}{\begin{equation}}
\newcommand{\eeq}{\end{equation}}
\newcommand{\bea}{\begin{eqnarray}}
\newcommand{\eea}{\end{eqnarray}}

\renewcommand{\[}{\left[}
\renewcommand{\]}{\right]}
\renewcommand{\(}{\left(}
\renewcommand{\)}{\right)}

\renewcommand{\S}{\mathcal{S}}
\renewcommand{\X}{\mathcal{X}}

\def\be{\begin{equation}}
\def\ee{\end{equation}}

\begin{document}

\begin{center}  
{\LARGE
\bf\color{blus}

A framework for missing-energy searches \\ 
with anomalous light vectors
}
\vspace{0.8cm}

{\bf 
Luca Di Luzio$^{a}$, 
Marco Nardecchia$^{b}$,
Stefano Scacco$^{b}$, 
Claudio Toni$^{c}$ }\\[5mm]

{\it $^a$Istituto Nazionale di Fisica Nucleare, Sezione di Padova, \\
Via F. Marzolo 8, 35131 Padova (PD), Italy}\\[1mm]
{\it $^b$Università degli Studi di Roma La Sapienza and INFN Section of Roma 1, \\
Piazzale Aldo Moro 5, 00185 Roma, Italy}\\[1mm]
{\it $^c$LAPTh, Université Savoie Mont-Blanc et CNRS,
74941 Annecy, France}

\vspace{0.3cm}
\begin{quote}
We study light spin-1 gauge bosons coupled to electroweak-anomalous currents. 
For generic charge assignments, anomaly cancellation requires new fermions (anomalons) that are chiral under the new abelian symmetry and carry electroweak charges. If their masses arise from the breaking of the new gauge symmetry, integrating them out generates Wess--Zumino interactions fixed by mixed-anomaly matching, providing the infrared description of the theory. 
We classify minimal anomalon spectra, derive the corresponding effective interactions, and combine experimental constraints with finite-naturalness considerations to bound the UV completion scale. 
Motivated by recent NA62 and Belle~II results, we then develop a unified phenomenological framework for the missing-energy signatures of these anomalous light vectors, focusing on scenarios where the new vector decays predominantly into neutrinos so that the leading probes are rare processes with invisible final states. 
As applications, we survey current and projected searches across flavour and electroweak observables, including $K\to\pi E_{\rm miss}$, $B\to K^{(*)}E_{\rm miss}$, and $Z\to\gamma E_{\rm miss}$, and discuss their interplay with direct searches for anomalons.
\end{quote}

\thispagestyle{empty}

\end{center}

\setcounter{tocdepth}{2}
\tableofcontents

\section{Introduction}
\label{sec:intro} 

The persistent absence of TeV-scale resonances has strengthened the case for exploring light and feebly interacting new states. Within this landscape, light spin-1 bosons are particularly well motivated. The standard benchmark is a secluded $\U(1)$ gauge boson kinetically mixed with the photon~\cite{Holdom:1985ag}, yielding universal couplings to the Standard Model (SM) electromagnetic current. While predictive, this setup can be too restrictive, and more general interactions of a light vector with SM fields are also worth considering. A theoretically motivated route beyond kinetic mixing is obtained by gauging the accidental global symmetries of the SM in the limit of massless neutrinos, namely baryon number $\U(1)_B$ and family lepton numbers $\U(1)_{L_i}$ ($i=e,\mu,\tau$). Within the SM field content, only the combinations $L_i-L_j$ are anomaly free~\cite{Foot:1990mn,He:1990pn,He:1991qd}. Hence, to gauge a general linear combination
\begin{equation}
\label{eq:genU1X_intro}
X \;=\; \alpha_B B \;+\; \sum_{i=e,\mu,\tau}\alpha_i L_i\,,
\end{equation}
one must introduce new fermions (``anomalons'') to cancel the $\U(1)_X$ anomalies, including mixed anomalies with $\SU(2)_L\times\U(1)_Y$. Under the assumption that the SM Yukawa operators are present at the renormalizable level, Eq.~\eqref{eq:genU1X_intro} is the most general gauged abelian symmetry of the SM (we also assume that the SM Higgs doublet is neutral under the new gauge symmetry). Barring the special cases $B/3-L_i$ and linear combinations thereof, anomaly cancellation requires anomalons charged under $\SU(2)_L\times \U(1)_Y$.

The anomalons $\Psi$ are chiral under the full gauge group and therefore cannot admit bare mass terms. Their masses must originate either from electroweak (EW) symmetry breaking or from the spontaneous breaking of $\U(1)_X$ via a new scalar $\S$ with vacuum expectation value (VEV) $v_X$. The first option is strongly constrained (for a recent analysis, see \cite{Barducci:2023zml}), so we focus on the second, in which $m_X\sim g_X v_X$ while $m_\Psi\sim y_\Psi v_X$, where $g_X$ and $y_\Psi$ are the gauge and Yukawa couplings of the new sector. In the regime of interest for a light vector, $g_X\ll 1$, a light $X_\mu$ can coexist with a large $v_X$, thereby pushing the ultraviolet (UV) completion (the anomalons) to $m_\Psi\gg m_X$ unless the Yukawas are exceedingly small. A central question is then how heavy this UV sector can be. Here we adopt the guiding perspective of ``finite naturalness''~\cite{Farina:2013mla}, and apply naturalness criteria to finite contributions to the Higgs mass. Requiring that the anomalons do not induce an excessively large finite correction to the Higgs mass yields an upper bound on $m_\Psi$, typically pointing to a multi-TeV scale for percent-level tuning. This provides a concrete target for the UV completion of light vectors coupled to anomalous SM currents.

If the anomalons are heavier than the EW scale, their impact on the low-energy physics of $X_\mu$ is captured by an effective field theory (EFT). Integrating out the anomalons generates dimension-4 Wess--Zumino (WZ) terms of the schematic form
$X (W\partial W + WWW)$ and $X B\partial B$,
which reproduce the anomalous variation of the effective action associated with the anomalous SM current coupled to $X_\mu$ (see e.g.~\cite{DHoker:1984izu,DHoker:1984mif,Preskill:1990fr,Feruglio:1992fp}). In the limit where anomalon masses originate from a SM-preserving VEV, the WZ coefficients are fully fixed by the requirement of canceling the $[\SU(2)_L^2\U(1)_X]$ and $[\U(1)_Y^2\U(1)_X]$ anomalies of the SM sector~\cite{Dror:2017nsg,DiLuzio:2022ziu}. Since WZ interactions exhibit an axion-like behaviour (equivalence theorem for the longitudinal mode of $X_\mu$), they induce amplitudes growing with energy~\cite{Dror:2017ehi,Dror:2017nsg}, leading to characteristic collider and flavor signatures. In particular, the $XW\partial W$ operator can source loop-induced flavor-violating processes, following a predictive flavor pattern which resembles the minimal flavor violation hypothesis~\cite{DAmbrosio:2002vsn}, while $XB\partial B$ contributes to $Z\to\gamma X$ at tree level (see e.g.~\cite{Ismail:2017fgq,Michaels:2020fzj,Davighi:2021oel,Kribs:2022gri}).

\clearpage

In this work we use this IR description to develop a general phenomenological framework for missing-energy searches for light vectors. Motivated in particular by the recent NA62 \cite{NA62:2024pjp,NA62:2025upx,LaThuile_2026_update} and Belle~II \cite{Belle-II:2023esi} results on rare meson decays with missing energy, we focus on scenarios where the dominant SM decay mode of the vector $X$ is into neutrinos, so that the leading probes are rare processes with invisible final states. These include radiative $Z$ decays and rare $K$, $D$, and $B$ transitions with missing energy, which together provide complementary sensitivity to the WZ interactions and to the underlying mixed-anomaly structure. We then specialize to \emph{vector} gauge currents of the form in Eq.~\eqref{eq:genU1X_intro} and to their minimal UV completions, complementing previous work where we analyzed light vectors coupled to \emph{chiral} currents~\cite{DiLuzio:2025qkc}. We classify the minimal anomalon spectra required by mixed-anomaly cancellation, derive the corresponding WZ coefficients in the EFT, and combine finite naturalness considerations with phenomenological constraints to delineate the viable parameter space.

A representative example within this broader framework is provided by 
models based on a gauged $\tau$-flavor symmetry: 
if the decay $X\to\tau^+\tau^-$ is kinematically forbidden, the vector can dominantly decay invisibly into neutrinos, yielding signals such as $B\to K^{(*)}X$~\cite{DiLuzio:2025qkc}. Such a setup could account for the $2.7\sigma$ excess reported by Belle~II in $B^+\to K^+E_{\rm miss}$~\cite{Belle-II:2023esi}. Ref.~\cite{DiLuzio:2025qkc} studied this possibility for light vectors coupled to chiral $\tau$-lepton currents and constructed explicit UV completions with $m_X\simeq 2.1~{\rm GeV}$~\cite{Bolton:2024egx} and a minimal anomalon sector, consistent with existing $e^+e^-\to\tau^+\tau^-X$ bounds and testable with future Belle~II data~\cite{Hoferichter:2025zjp}. Here, we instead focus on the more minimal and theoretically motivated case of vector currents, for which the SM Yukawa operators can be written at the renormalizable level, and we emphasize the resulting missing-energy phenomenology, directly relevant for Belle~II in $B\to K^{(*)}E_{\rm miss}$ and for NA62/KOTO in $K\to\pi E_{\rm miss}$.

The paper is organized as follows. Section~\ref{sec:classification} lays out the anomaly-cancellation conditions implied by gauging \eq{eq:genU1X_intro} and the model-selection criteria that will guide our UV constructions. Section~\ref{sec:missen}, instead, 
is self-contained and focuses on IR physics:\footnote{Readers focused on low-energy phenomenology and missing-energy signatures may skip the general discussion (Section~\ref{sec:classification}) and explicit models (Section~\ref{sec:UVmodel}), and go straight to Section~\ref{sec:missen}.}  
it develops the missing-energy framework and surveys the leading bounds from $Z$, $K$, $D$, and $B$ decays.
In Section~\ref{sec:UVmodel} we exhibit representative UV completions and discuss current constraints on anomalons from 
direct searches. 
Our conclusions are collected in Section~\ref{sec:conclusion}. Appendix~\ref{App:anomalonsetups} provides technical details on the classification of anomalon setups.

\section{Anomaly cancellation and model selection}
\label{sec:classification}

In addition to the SM fields, a setup of anomaly-canceling fermions as in 
\Table{tab:genU1fieldcontent} is required
for the gauging of Eq.~\eqref{eq:genU1X_intro}. On top of the anomalon
fields $\Psi_\alpha$ ($\alpha = 1, \ldots, N_\Psi$),
we also extend
the scalar sector of the SM in order to spontaneously break the $\U(1)_X$ symmetry. 
Note that we have included $N_\N$ copies of chiral SM-singlet fermions $\N_{\beta}$ 
($\beta = 1, \ldots, N_\N$)
which play a role for the cancellation of the 
$[\U(1)_X]$ and $[\U(1)^3_X]$ anomalies, 
but whose presence does not impact the calculation of the WZ terms in the EW sector. For simplicity, we take the anomalon fields to be color singlets, since the SM accidental symmetries in Eq.~\eqref{eq:genU1X_intro} yield no $[\SU(3)_C^2\U(1)_X]$ anomaly. 
The generalization to colored anomalon fields is discussed 
in Appendix~\ref{app:gencolanom}, where it is shown that no new relevant solutions are obtained. 

\begin{table}[h!]
	\centering
	\begin{tabular}{|c|c|c|c|c|c|}
	\hline
	Field & Lorentz & $\SU(3)_C$ & $\SU(2)_L$ & $\U(1)_Y$ & $\U(1)_{X}$ \\ 
	\hline 
	$q^{i}_L$ & $(\tfrac{1}{2},0)$ & 3 & 2 & 1/6 & $\alpha_B/3$ \\ 
	$u^{i}_R$ & $(0,\tfrac{1}{2})$ & 3 & 1 & 2/3 & $\alpha_B/3$ \\ 
	$d^{i}_R$ & $(0,\tfrac{1}{2})$ & 3 & 1 & $-1/3$ & $\alpha_B/3$ \\ 
	$\ell^{i}_L$ & $(\tfrac{1}{2},0)$ & 1 & 2 & $-1/2$ & $\alpha_i$ \\ 
	$e^{i}_R$ & $(0,\tfrac{1}{2})$ & 1 & 1 & $-1$ & $\alpha_i$ \\ 
	$H$ & $(0,0)$ & 1 & 2 & 1/2 & $0$ \\
	\hline 
	\rowcolor{cyan!20} 
	$\Psi_\alpha$ & $(\tfrac{1}{2},0)$ & 1 & $n_\alpha$ & $\Y_{\alpha}$ & $\X_{\alpha}$ \\
	\rowcolor{cyan!20} 
	$\N_{\beta}$ & $(0,\tfrac{1}{2})$ & 1 & 1 & $0$ & $X_{\beta}$ \\
	\hline
    $\S$ & $(0,0)$ & 1 & 1 & 0 & $\X_{\S}$ \\  
	\hline
	\end{tabular}	
	\caption{\label{tab:genU1fieldcontent} 
	Anomaly-free field content for an $\SU(3)_C \times \SU(2)_L \times \U(1)_Y \times \U(1)_X$ gauge theory, 
	with $X = \alpha_B B + \sum_{i=e,\mu,\tau} \alpha_i L_i$. 
    The SM field content is augmented by a set of 
    anomalon fields (highlighted in blue) 
    and a new complex scalar. 
    The conditions on the $\U(1)_X$ charges fulfilling the cancellation of gauge anomalies 
	are reported in Eqs.~\eqref{eq:anom1}--\eqref{eq:anom5}.
    }
\end{table}
The schematic Lagrangian associated with the field content in Table~\ref{tab:genU1fieldcontent} can be written as
(besides the SM, kinetic terms, and the $\U(1)_X$-breaking scalar potential discussed in Section \ref{sec:UVmodel}$)$
\be
\lag \;\supset\; \lag_{\Psi\S}^\text{Yuk} \;+\; \lag_{\ell\Psi} \;+\; \lag_{\Psi H}^\text{Yuk} \;+\; \lag_\N \, .
\ee
The individual contributions have the following roles. The anomalon masses are generated by $\lag_{\Psi\S}^\text{Yuk}$ after spontaneous $\U(1)_X$ breaking. Possible mixing between anomalons and SM fields can proceed via SM leptons or the Higgs doublet. In the class of models we explicitly classify below, the $\U(1)_X$ charge assignments are chosen such that 
mixings between SM leptons and anomalons are absent, i.e.~$\lag_{\ell\Psi}=0$. Some models can nevertheless feature Higgs-induced off-diagonal Yukawa interactions between different anomalon multiplets, encoded in $\lag_{\Psi H}^\text{Yuk}$: these terms enable decays within the anomalon sector and can generate tree-level mass splittings among the $\SU(2)_L$ components of a given multiplet. 
Finally, $\lag_\N$ denotes the SM-singlet sector. It includes singlet mass terms (either Majorana masses or Yukawa couplings to $\S$) and, depending on the singlet field content and charge assignments, possible mixings with SM leptons and/or anomalons. Since these features are model-dependent and typically have limited impact on the mixed-anomaly/WZ phenomenology, we do not attempt a general classification; representative examples are discussed in Section \ref{sec:UVmodel}.

To ensure gauge-anomaly cancellation, the following five conditions must be satisfied:
\begin{align}
\label{eq:anom1}
[\U(1)_X] &:\quad \sum_\alpha n_\alpha \X_{\alpha} - \sum_{\beta} X_{\beta} + \alpha_{e} + \alpha_{\mu} + \alpha_{\tau}  = 0 \, , \\
\label{eq:anom2}
[\U(1)^3_X] &:\quad \sum_\alpha n_\alpha \X^3_{\alpha} - \sum_{\beta} X^3_{\beta} + \alpha^3_{e} + \alpha^3_{\mu} + \alpha^3_{\tau}  = 0 \, , \\
\label{eq:anom3}
[\SU(2)^2_L \U(1)_X] &:\quad \sum_\alpha \frac{1}{12} n_\alpha(n_\alpha^2 -1) \X_{\alpha} + \frac{3}{2}\alpha_{B+L} = 0 \, , \\ 
\label{eq:anom4}
[\U(1)^2_Y \U(1)_X] &:\quad \sum_\alpha n_\alpha \Y_{\alpha}^2 \X_{\alpha}   
- \frac{3}{2}\alpha_{B+L} = 0 \, , \\
\label{eq:anom5}
[\U(1)_Y \U(1)^2_X] &:\quad  \sum_\alpha n_\alpha \Y_{\alpha} \X_{\alpha}^2
= 0 \, ,
\end{align}
where we introduced the parameter
\be
\label{eq:BL}
\alpha_{B+L} \equiv \alpha_B + \frac{1}{3} ( \alpha_{e} + \alpha_{\mu} + \alpha_{\tau} ) \, .
\ee
An additional condition comes from the non-perturbative Witten anomaly for the $\SU(2)_L$ gauge group which requires an even number of Weyl fermions in the representation $n = 4l + 2$ with $l\in \mathbb{N}$~\cite{Witten:1982fp,Wang:2018qoy}.

Remarkably, the cancellation of mixed gauge anomalies Eqs.~\eqref{eq:anom3}--\eqref{eq:anom5} depends solely on the combination in Eq.~\eqref{eq:BL}, independent of the specific quantum numbers of the SM fields. This renders the proposed classification broadly applicable to any scenario featuring an anomalous low-energy current, assuming only the presence of SM Yukawa operators at the renormalizable level.

Among the general solutions of Eqs.~\eqref{eq:anom1}--\eqref{eq:anom5}, we are interested on those which 
satisfy the finite naturalness criterion 
and are phenomenologically viable, as discussed in Section~\ref{subsec:finitenaturalness} and~\ref{subsec:phenomenology}.

\subsection{Finite naturalness criterion}
\label{subsec:finitenaturalness}
By finite naturalness \cite{Farina:2013mla}, we refer to the criterion where uncalculable quadratic divergences to the Higgs mass are ignored, and naturalness constraints are applied solely to the calculable, finite corrections induced by new heavy particles in a given extension of the SM.
Therefore, in the regime of interest where the new vector boson is lighter than the EW scale, 
only anomalons would contribute to the Higgs mass correction.  
We then constrain their contribution, and thus their mass, by imposing an upper limit on the Higgs-mass fine-tuning parameter 
\be
\Delta = \frac{\delta m_h^2}{M_h^2} \lesssim 100 \, ,
\ee
where $m_h$ is the Higgs mass parameter appearing in the Lagrangian, and $M_h$ is the pole Higgs mass. This corresponds to a fine tuning of no more than 1\%, which approximately mirrors the fine tuning on supersymmetric models in response to LHC direct searches (see e.g.~\cite{vanBeekveld:2019tqp}). 

Following \cite{Farina:2013mla}, the two-loop correction due to 
a new fermion 
with SM quantum numbers $(1,n,\Y)$ is
\be \label{eq:naturalness2loop}
\delta m_h^2 = \frac{Cn M_\Psi^2}{(4\pi)^4}\left(\frac{n^2-1}{4} g_2^4 + \Y^2 g_Y^4\right)\left(6 \log\frac{M_\Psi^2}{\mu^2}-1\right) \, ,
\ee
where $M_\Psi$ is the anomalon mass scale and $\mu$ is the renormalization scale, while $C = 1 (2) $ for Majorana (Dirac) fermions.

Furthermore, some models also feature anomalon Yukawa terms with the SM Higgs, which are model-dependent and yield a correction $\delta m_h^2$ already at one loop. However, bounds from EW precision observables and Higgs couplings strongly constrain the possibility of chiral anomalons~\cite{Barducci:2023zml}; accordingly, such contributions are expected to be subdominant and will be neglected in the following.

As a result of the criterion $\Delta \lesssim 100$, the anomalon mass scale cannot be arbitrarily 
decoupled from the TeV scale.

\subsection{Phenomenological constraints}
\label{subsec:phenomenology}
The phenomenological selection is based on the following criteria:
\begin{itemize}

\item[(i)] No massless anomalon fermion must be left after the spontaneous symmetry breaking (SSB) of gauge symmetries.
Furthermore, in order to avoid bounds from the SM Higgs sector, which highly disfavours chiral anomalons~\cite{Barducci:2023zml}, we require the anomalon fields to take mass mostly from the $\S$ boson after the SSB of the new symmetry.

The mass term can be either Majorana-like (ML) $\bar\Psi_\alpha \Psi_\alpha^C \langle \S^{(\dagger)} \rangle$, where $C$ denotes the charge conjugation, or Dirac-like (DL) $\bar\Psi_{\alpha_1} \Psi_{\alpha_2}^C\langle \S^{(\dagger)} \rangle$. 
The former case requires $\Y_\alpha=0$ and $n_\alpha$ odd,\footnote{Majorana mass terms with a pseudoreal representation under $\SU(2)_L$ identically vanish.} with $n_\alpha>1$ otherwise the field is a SM-singlet, and $\X_\alpha=\pm\X_\S/2$; while the latter requires an anomalon pair such that $n_{\alpha_1}=n_{\alpha_2}$, $\Y_{\alpha_1}=-\Y_{\alpha_2}(\neq0)$ and $\X_{\alpha_1}+\X_{\alpha_2}=\pm\X_\S$.
Hence, we are looking for solutions of Eqs.~\eqref{eq:anom1}--\eqref{eq:anom5} composed by ML fields or DL pairs, which are both SM anomaly-free.

Note that Eq.~\eqref{eq:anom4} requires at least a DL pair to be solved.

\item[(ii)] 

Stable anomalon states are required to be electrically neutral.

Charged stable particles are tightly constrained by anomalous energy loss and charged-track searches at the LHC (see e.g.~\cite{ATLAS:2023zxo}), which typically push their masses above the TeV scale, and by cosmology, since the relic abundance of stable charged relics is severely limited. Given the strength of these bounds, the mass range compatible with finite naturalness is either excluded or, at best, restricted to a narrow window around the multi-TeV scale. We therefore require any stable anomalon state to be neutral.

If mixing terms between anomalons and SM fermions (or additional SM-singlets) are absent, the lightest particle (LP) in the anomalon sector is stable\footnote{In this case the anomalon sector enjoys an accidental $Z_2$ symmetry, $\Psi_\alpha\to-\Psi_\alpha$, which forbids the decay of the LP in the anomalon multiplet.} and must consequently be electrically neutral.

In the cases where anomalon Yukawa terms with the SM Higgs boson are absent, this requirement holds for each single non-degenerate ML field and DL pair. In order to contain a neutral component, $n_\alpha$ must be odd for a ML field, while $\Y_\alpha\in[-(n_\alpha-1)/2, \dots,(n_\alpha-1)/2]$ for a DL pair.
The mass splitting among the neutral and charged components is given by electroweak radiative corrections, estimated  as~\cite{Cirelli:2005uq}
\be
\label{eq:Deltam}
m_{Q} - m_0 \approx 166 \, \text{MeV} \times \[ Q \(Q + \frac{2 \Y_\alpha}{c_W} \) \] \, ,
\ee
where $c_W$ is the cosine of the weak angle.
The LP is neutral for a DL pair if and only if $|\Y_\alpha|=(n_\alpha-1)/2$ (see Ref.~\cite{DiLuzio:2025qkc} for details),  while this is always true for a ML field.

In the cases where anomalon Yukawa terms with the SM Higgs boson are present, decays between different multiplets are possible and the mass splitting among the neutral state and the charged ones depends on the details of the anomalon Yukawa sector.

\item[(iii)] The anomalon $\SU(2)_L$ representation index $n_\alpha$ must not spoil the perturbativity of the interacting theory.
The pertubativity bound can be imposed on the diagonal component of $\SU(2)_L$, with generator $T_{W_3}$, as
\be
\frac{\left(T_{W_3} g\right)^2}{4\pi}\lesssim 1 \ ,
\ee
for each component of the multiplet and where $g\simeq 0.65$ is the $\SU(2)_L$ gauge coupling. The strongest bound comes from the largest diagonal isospin value $|T_{W_3}|=(n_\alpha-1)/2$, for which $n_\alpha\lesssim10$. 

\item[(iv)] The present discussion focuses mostly on classifying the solutions of Eqs.~\eqref{eq:anom3}--\eqref{eq:anom5}, since Eqs.~\eqref{eq:anom1}--\eqref{eq:anom2} can be solved just by introducing SM-singlets, whose presence does not affect the calculation of the WZ terms in the EW sector. Note also that Eqs.~\eqref{eq:anom3}--\eqref{eq:anom5} depend linearly on the $B+L$ projection of Eq.~\eqref{eq:genU1X_intro}, so that the anomalon solutions depends only on $\alpha_{B+L}$. Instead Eqs.~\eqref{eq:anom1}--\eqref{eq:anom2} have a non-linear dependence on the lepton family number charges, which makes it difficult to obtain a general solution. The choice of the SM-singlet field content, potentially related to neutrino mass generation, is then model-dependent and will not be considered in the classification.

\end{itemize}

Altogether, these criteria, along with constraints from direct collider searches (discussed in Section~\ref{sec:UVmodel}) and finite-naturalness considerations, confine the anomalon mass scale to the range from a few hundred GeV to a few TeV.

\subsection{Minimal anomalon spectra}
\label{sec:minanomspect}

We now present solutions to the anomaly-cancellation conditions in Eqs.~\eqref{eq:anom3}--\eqref{eq:anom5}, for $N_\Psi\leq 4$, that satisfy our naturalness and phenomenological criteria. The systematic classification can be found in Appendix~\ref{App:anomalonsetups}. 
As multiplets get larger, the naturalness upper limit on anomalon scale decreases, until eventually the model is  excluded by collider searches, which is why in our classification only the simplest models are explicitly shown.

Viable solutions involve $N_\Psi \geq 3$. 
For $N_\Psi = 3$ one has a ML field and a DL pair, 
denoted respectively as  
\be
\label{eq:fieldsNeq3}
\Psi_1\sim\left(n, 0,\X_1\right) \, , \quad
\Psi_2\sim\left(m, \Y,\X_2\right) \, , \quad
\Psi_3\sim\left(m, -\Y,\X_3\right) \, , 
\ee
under $\SU(2)_L\times \U(1)_Y\times \U(1)_X$. 
Valid solutions that satisfy all criteria require $-\X_1 = \X_2 = \X_3 = \X_\S /2$ (see Appendix~\ref{subApp:N=3}), and are listed in Table~\ref{tab:sol1ML1DL}.
\begin{table}[ht!]
\centering
\begin{tabular}{|c|c|c|c|c|}
\hline
$n$ & $m$ & $\Y$ & $\X_\S/(3\alpha_{B+L})$ & Naturalness (TeV) \\
\hline
\rowcolor{piggypink}
$3$ & $2$ & $1/2$ & $1$ & $7.3\times \sqrt{\Delta/100}$\\
\rowcolor{piggypink}
$5$ & $3$ & $1$ & $1/6$ & $3.4\times \sqrt{\Delta/100}$\\
$7$ & $4$ & $3/2$ & $1/18$ & $2.0\times \sqrt{\Delta/100}$\\
$9$ & $5$ & $2$ & $1/40$ & $1.4\times \sqrt{\Delta/100}$\\
$11$ & $6$ & $5/2$ & $1/75$ & $1.0\times \sqrt{\Delta/100}$ \\
\hline
\end{tabular}
\caption{List of all the models with one ML and one DL pair in \eq{eq:fieldsNeq3} satisfying the anomaly-cancellation conditions in Eqs.~\eqref{eq:anom3}--\eqref{eq:anom5}, with naturalness bound above 1 TeV assuming $\Delta = 100$. The last column reports the finite-naturalness bound obtained from Eq.~\eqref{eq:naturalness2loop}. The models highlighted in pink are analyzed in Section~\ref{sec:1DL1MLmodel}.}
\label{tab:sol1ML1DL}
\end{table}
$\X_\S$ values are crucial because, as shown in Eq.~\eqref{eq:massX}, smaller $\X_\S$ corresponds to larger $\U(1)_X$ breaking scale, at fixed mediator mass $m_X$ and coupling $g_X$.

For $N_\Psi = 4$, either two DL pairs, or one DL pair and two ML fields are possible. For the latter case, no acceptable model is found. Thus, we consider two DL pairs defined as
\be
\label{eq:fieldsNeq4}
\begin{split}
\Psi_1\sim\left(n, \Y_1,\X_1\right) \, ,
&\quad
\Psi_2\sim\left(n, -\Y_1,\X_2\right) \, , \\
\Psi_3\sim\left(m, \Y_3,\X_3\right) \, ,
&\quad
\Psi_4\sim\left(m, -\Y_3,\X_4\right) \, .
\end{split}
\ee
Valid solutions require $\X_1 = -\X_4$ and $\X_2 = -\X_3$ with $\X_S=\X_1+\X_2$
(see Appendix~\ref{subApp:N=4}), and are shown in Table~\ref{tab:sol2DL}.
\begin{table}[ht!]
\centering
\begin{tabular}{|c|c|c|c|c|c|c|}
\hline
$n$ & $m$ & $\Y_1$ & $\Y_3$ & $\X_1/(3\alpha_{B+L})$ & $\X_2/(3\alpha_{B+L})$ & Naturalness (TeV) \\
\hline
\rowcolor{piggypink}
$1$ & $2$ & $1$ & $1/2$ & $\X-1/2$ & $-\X-1/2$ & $12.2\times \sqrt{\Delta/100}$\\
$2$ & $3$ & $3/2$ & $1$ & $\X-1/6$ & $-\X-1/6$ & $5.4\times \sqrt{\Delta/100}$\\
$3$ & $4$ & $2$ & $3/2$ & $\X-1/12$ & $-\X-1/12$ & $3.2\times \sqrt{\Delta/100}$\\
$4$ & $5$ & $5/2$ & $2$ & $\X-1/20$ & $-\X-1/20$ & $2.2\times \sqrt{\Delta/100}$\\
$5$ & $6$ & $3$ & $5/2$ & $\X-1/30$ & $-\X-1/30$ & $1.6\times \sqrt{\Delta/100}$\\
$6$ & $7$ & $7/2$ & $3$ & $\X-1/42$ & $-\X-1/42$ & $1.2\times \sqrt{\Delta/100}$\\
\hline
\end{tabular}
\caption{List of all the models with one ML and one DL pair in \eq{eq:fieldsNeq4} satisfying the anomaly-cancellation conditions in Eqs.~\eqref{eq:anom3}--\eqref{eq:anom5}, with naturalness bound above 1 TeV assuming $\Delta = 100$. The last column reports the finite-naturalness bound obtained from Eq.~\eqref{eq:naturalness2loop}. The model highlighted in pink is analyzed in Section~\ref{sec:1DL1MLmodel}.}
\label{tab:sol2DL}
\end{table}
We remark that these solutions exhibit a degeneracy in the $\X$ charge under an arbitrary global shift,  
which can be fixed so as to forbid anomalon-SM fermion mixing, either via Dirac mass terms or via Yukawa operators involving the Higgs field. The specific realization $\X = 1/6$, for the first row in Table~\ref{tab:sol2DL} has been studied in the context of the phenomenology of gauged $\U(1)_B$ in \cite{Michaels:2020fzj,Armbruster:2025jqs} and with a focus on dark matter in \cite{Ellis:2017tkh}.

The naturalness bounds shown for $N_\Psi=3,4$ do not include the one-loop contribution from anomalon Yukawa interactions with the SM Higgs, which, when present, depends on the specific couplings. The resulting estimate is therefore conservative.

Finally, we have checked that allowing for colored anomalon fields does not yield additional phenomenologically relevant solutions for $N_\Psi\leq 4$; see Appendix~\ref{App:anomalonsetups}.

\section{Missing-energy signatures of light vectors}
\label{sec:missen}

In this section we develop a general phenomenological framework for missing-energy searches of light vectors coupled to electroweak-anomalous currents. In the mass range of interest, the dominant SM decay mode of these vectors can be into neutrinos, so that the leading probes are rare processes with invisible final states. In particular, the WZ interactions generated by integrating out the electroweak anomalons induce characteristic signals in radiative $Z$ decays and in rare meson transitions with missing energy, providing complementary sensitivity across the $K$, $D$, and $B$ sectors.

We begin with an EFT description of the relevant low-energy couplings in terms of WZ operators, keeping the discussion as model independent as possible. We then review the main experimental searches for $E_{\rm miss}$ signatures in $Z$, $K$, $D$, and $B$ decays, and summarize the corresponding constraints and projections. As a representative example within this broader framework, we discuss the gauged $\tau$-flavor symmetry benchmark (corresponding to $\alpha_B=\alpha_e=\alpha_\mu=0$ and $\alpha_\tau=1$ in Eq.~\eqref{eq:genU1X_intro}), motivated by the fact that a $\tau$-philic vector can dominantly decay invisibly into neutrinos when $X\to\tau^+\tau^-$ is kinematically forbidden,\footnote{Notably, a vector boson with visible decay modes, such as $X\to e^+e^-$ or $X\to \mu^+\mu^-$, may still lead to a missing-energy signature if it is sufficiently long-lived to decay outside the detector, or if it decays predominantly into a light, stable, neutral fermion of the anomalon sector. The former case, however, requires couplings so small that the corresponding bounds are essentially evaded throughout almost all of parameter space, unless the visible decays are kinematically forbidden, in which case $X$ is trivially invisible. The latter case is model-dependent and would need to be assessed separately in each realization.} yielding signals such as $B\to K^{(*)}X$~\cite{DiLuzio:2025qkc}. This setup has been considered as a possible explanation of the $2.7\sigma$ excess reported by Belle~II in $B^+\to K^+E_{\rm miss}$~\cite{Belle-II:2023esi}, and explicit UV completions were constructed in Ref.~\cite{DiLuzio:2025qkc} for light vectors coupled to chiral currents. Here we instead focus on the more minimal case of vector gauge currents in Eq.~\eqref{eq:genU1X_intro} and use the $\tau$-flavor benchmark to illustrate the interplay between missing-energy searches and the IR WZ description.

The connection to explicit UV completions and to direct searches for anomalons is discussed in Section~\ref{sec:UVmodel}, where we build representative models based on the anomaly-canceling spectra identified in Section~\ref{sec:classification}.

\subsection{EFT and Wess-Zumino terms} 
\label{sec:EFTWZ}

The EW anomalies are 
cured in the UV with the presence of anomalon fields,  
which are chiral under the $\U(1)_X$ 
and transform non-trivially under the EW group.
In the low-energy EFT, valid below the $\U(1)_X$-breaking scale, the EW anomalies are compensated by effective operators which appear upon integrating out the anomalon fields at one loop. 
Assuming the phenomenologically motivated case where 
the mass of the anomalons mostly stems from the VEV of a SM-singlet field,
one finds \cite{DHoker:1984izu,DHoker:1984mif,Dror:2017ehi,Dror:2017nsg,DiLuzio:2022ziu} 
\begin{align}
\label{EFT}
\lag_{\rm X}
\supset &-g_X
J^\mu_{X,\text{SM}}
X_\mu
-g_{X}g'^{2}\frac{{\cal A}_{XYY}^\text{SM}}{24\pi^{2}}\epsilon^{\alpha\mu\nu\beta}X_{\alpha}B_{\mu}\partial_{\beta}B_{\nu} \nonumber \\
&-g_{X}g^{2}\frac{{\cal A}_{XWW}^\text{SM}}{24\pi^{2}}\epsilon^{\alpha\mu\nu\beta}X_{\alpha}W_{\mu}^{a}\partial_{\beta}W_{\nu}^{a}
+\text{non-abelian terms}
\,,  
\end{align}
where $a = 1,2,3$ and we neglected non-abelian $W$ terms scaling with an extra gauge coupling $g$.  
The anoumalous traces are defined as
\beq
{\cal A}_{\alpha\beta\gamma}\equiv\text{Tr}\left[T_\alpha \{T_\beta,T_\gamma\}\right]\Big|_\text{R} - \text{Tr}\left[T_\alpha \{T_\beta,T_\gamma\}\right]\Big|_\text{L} \, , 
\eeq 
with $T_{\alpha}$ being any generator of the gauge group.  
Specifying the latter to the SM anomalous traces of 
\eq{EFT}, we have
${\cal A}_{XYY}^\text{SM}=-{\cal A}_{XWW}^\text{SM}=3\alpha_{B+L}$, with the $X$ generator defined in Eq.~\eqref{eq:genU1X_intro}. The divergence of the SM current coupled to $X$ is given by
\begin{align}
\partial_\mu 
J_{X,\text{SM}}^\mu &= g'^2\frac{{\cal A}_{XYY}^\text{SM}}{48\pi^{2}} \epsilon^{\alpha\mu\beta\nu}(\partial_{\alpha}B_{\mu})(\partial_{\beta}B_{\nu}) +g^2\frac{{\cal A}_{XWW}^\text{SM}}{48\pi^{2}} \epsilon^{\alpha\mu\beta\nu}(\partial_{\alpha}W^a_{\mu})(\partial_{\beta}W^a_{\nu}) \nonumber \\
&+ \text{non-abelian terms} \, .
\end{align}
In the gauge-less limit, i.e.~$g_X\to0$ while $m_X/g_X$ fixed,
the longitudinal mode of the $X$ field dominates the physical amplitudes. Through the Goldstone boson equivalence theorem, we can isolate the longitudinal mode in this limit via the substitution $X_\mu\to\partial_\mu\xi/m_X$, 
where $\xi$ denotes the Goldstone field. Applying this to \eq{EFT} and integrating by parts, leads to the axion-like operator
\begin{equation}
\label{eq:LXplusLWZ}
{\cal L}_X \supset 
g_{X}g^{2}\frac{{\cal A}_{XWW}^\text{SM}}{16\pi^{2}}\frac{\xi}{m_X} \, \epsilon^{\alpha\mu\beta\nu}(\partial_{\alpha}W^a_{\mu})(\partial_{\beta}W^a_{\nu}) \, .
\end{equation}
Upon integrating out the $W$ boson at the one-loop level, 
the axion-like operator $\xi W^{-}\tilde{W}^{+}$ yields the effective interaction 
\beq
\label{EFT FCNC}
g_{\xi d_{i}d_{j}}\bar{d}_{j}\gamma^{\mu}P_{L}d_{i}\,(\partial_{\mu}\xi/m_{X})\,+g_{\xi u_{i}u_{j}}\bar{u}_{j}\gamma^{\mu}P_{L}u_{i}\,(\partial_{\mu}\xi/m_{X})\,+\text{h.c.} \, , 
\eeq
in terms of the effective couplings \cite{Izaguirre:2016dfi,Sun:2025hep}
\begin{align}
\label{eq:gxididj}
g_{\xi d_{i}d_{j}}&=-\frac{3g_{X}g^{4}}{(4\pi)^{4}}{\cal A}_{XWW}^\text{SM}\sum_{\alpha=u,c,t}V_{\alpha i}V_{\alpha j}^{*}F(m_{\alpha}^{2}/m_{W}^{2}) \, , \\
\label{eq:gxiuiuj}
g_{\xi u_{i}u_{j}}&=-\frac{3g_{X}g^{4}}{(4\pi)^{4}}{\cal A}_{XWW}^\text{SM}\sum_{\beta=d,s,b}V^*_{i\beta}V_{j\beta}F(m_{\beta}^{2}/m_{W}^{2}) \, ,
\end{align}
with $V$ denoting the CKM matrix and the loop function
\beq 
F(x)=\frac{x(1+x(\ln x-1))}{(1-x)^{2}} \, .
\eeq
This gives rise to rare meson decays induced by the quark-flavor transitions $d_i\to d_j X$ and $c\to u X$ (cf.~Table~\ref{tab:mesondecays}), which follow a 
predictive minimal-flavor-violation pattern; 
and $X$ appearing as missing energy when its dominant decay mode is into neutrinos. 

The theoretical expressions of the flavor-violating meson decay rates are given by
\begin{align}
\label{eq:BRpseudo}
\Gamma(P(q_j)\to \tilde{P}(q_i) X)&=\frac{1}{32\pi}|g_{\xi q_i q_j}|^2 |\vec{k}| \frac{m_P^2}{m_X^2} \left(1-\frac{m_{\tilde{P}}^2}{m_P^2}\right)^2 f_0^2(m_X^2) \ , \\
\Gamma(P(q_j)\to V(q_i) X)&=\frac{1}{8\pi}|g_{\xi q_i q_j}|^2 \frac{|\vec{k}|^3}{m_X^2} A_0^2(m_X^2) \ , 
\end{align}
with $|\vec{k}|=\lambda^{1/2}(m_P^2,m_{\tilde{P},V}^2,m_X^2) / (2 m_P)$ and $\lambda(a,b,c) = a^2+b^2+c^2-2ab-2ac-2bc$. Here, $P(\tilde{P})$ and $V$ denote respectively pseudoscalar and vector mesons, while the meson form factors are defined as
\begin{align}
\label{eq:BRpseudo2}
\braket{\tilde{P}(q_i)|\overline{q}_i q_j|P(q_j)}&=\frac{m_P^2-m_{\tilde{P}}^2}{m_j-m_i}f_0(q^2) \ , \\
\braket{V(q_i)|\overline{q}_i\gamma_5 q_j|P(q_j)}&=-\frac{2im_V (\epsilon^*\cdot q)}{m_j+m_i}A_0(q^2) \ , 
\end{align}
with $q=p_P-p_{\tilde{P},V}$.\footnote{Form factors are taken as follows. $B\to K$: $f_0(q^2)$ from Ref.~\cite{Gubernari:2023puw}, identical for charged/neutral modes neglecting isospin breaking. $B\to K^{*}$: $A_0(q^2)$ from Ref.~\cite{Gubernari:2023puw}, again identical for charged/neutral modes. $B\to\pi$: $f_0(q^2)$ for $B^0\to\pi^-$ from Ref.~\cite{Biswas:2022yvh}, other channels related by isospin. $B\to\rho$: $A_0(q^2)$ for $B^0\to\rho^-$ from Ref.~\cite{Biswas:2022yvh}, other channels related by isospin. $K\to\pi$: $f_0(q^2)$ for $K^0\to\pi^-$ from Ref.~\cite{Carrasco:2016kpy}, other channels related by isospin. $D\to\pi$: $f_0(q^2)$ for $\overline{D}^0\to\pi^+$ from Ref.~\cite{Lubicz:2017syv}, other channels related by isospin.} 
For the decay $K_L\to\pi^0 X$, one needs to replace $g_{\xi q_i q_j} \to \sqrt{2}~\text{Im}(g_{\xi ds})$ in Eq.~\eqref{eq:BRpseudo}.

Similarly to \eq{eq:LXplusLWZ}, for the neutral EW sector 
one obtains the operator
\begin{align}
\label{eq:xidZdAlag}
&{\cal L}_X 
\supset 
c_{XZ\gamma} 
\frac{\xi}{m_{X}}\epsilon^{\alpha\mu\beta\nu}(\partial_{\alpha}Z_{\mu}^{a})(\partial_{\beta}A_{\nu}) \, , 
&c_{XZ\gamma}=g_{X}gg'\frac{c_W^2{\cal A}_{XWW}^\text{SM}-s_W^2{\cal A}_{XYY}^\text{SM}}{8\pi^{2}}  \, ,
\end{align}
with $s_W
=g'/\sqrt{g^2+g'^2}$,
which induces the decay $Z\to\gamma X$ with a rate given by \cite{Dror:2017ehi,Michaels:2020fzj}
\begin{align}
\Gamma(Z\to\gamma X)&=\frac{|c_{XZ\gamma}|^2}{96 \pi}\frac{(m_Z^2-m_X^2)^2}{m_Z m_X^2}\left(1-\frac{m_X^4}{m_Z^4}\right) \, .
\end{align}

\begin{table}[t!]
\centering
\begin{tabular}{c|c|c}
Flavor transition & Meson decay & Experiment \\
\midrule
\midrule
\multirow{2}{*}{$s \to d$} 
  & $K^+ \to \pi^+ E_{\rm miss}$ & NA62~\cite{NA62:2024pjp} \\
  & $K_L \to \pi^0 E_{\rm miss}$ & KOTO~\cite{KOTO:2024zbl} \\
\midrule
\midrule
\multirow{1}{*}{$c \to u$} 
  & $D^+ \to \pi^+ E_{\rm miss}$ & CLEO~\cite{CLEO:2008ffk} \\
\midrule
\midrule
\multirow{4}{*}{$b \to d$} 
  & $B^+ \to \pi^+ E_{\rm miss}$ & Belle~\cite{Belle:2013tnz} \\
  & $B^0 \to \pi^0 E_{\rm miss}$ & Belle~\cite{Belle:2013tnz} \\
  & $B^+ \to \rho^{+} E_{\rm miss}$ & Belle~\cite{Belle:2013tnz} \\
  & $B^0 \to \rho^{0} E_{\rm miss}$ & Belle~\cite{Belle:2013tnz} \\ 
\midrule
\midrule
\multirow{4}{*}{$b \to s$} 
  & $B^+ \to K^+ E_{\rm miss}$ & Belle~II~\cite{Belle-II:2023esi}, BaBar~\cite{BaBar:2013npw} \\
  & $B^0 \to K^0 E_{\rm miss}$ & BaBar~\cite{BaBar:2013npw} \\
  & $B^+ \to K^{*+} E_{\rm miss}$ & BaBar~\cite{BaBar:2013npw} \\
  & $B^0 \to K^{*0} E_{\rm miss}$ & BaBar~\cite{BaBar:2013npw} \\
\end{tabular}
\caption{List of flavor-violating meson decays with missing-energy signatures, and related experimental searches. 
}
\label{tab:mesondecays}
\end{table}

\subsection{Missing-energy searches}

The missing-energy phenomenology of this framework is governed by the EW probe $Z\to\gamma E_{\rm miss}$ and by rare meson decays associated with $d_i\to d_j E_{\rm miss}$ and $c\to u E_{\rm miss}$ transitions. The corresponding observables are summarized in Table~\ref{tab:mesondecays} and discussed below.

\subsubsection{$Z \to \gamma E_{\rm miss}$}
\label{sec:Zdecay}

The strongest constraint on these decays is provided by the L3 search at LEP, which yields~\cite{L3:1997exg}
\be
\label{eq:LEPbound}
{\mathcal B}(Z\to\gamma X) \lesssim 10^{-6} \text{  at 95\% C.L.} \, ,
\ee
based on an integrated luminosity $L_{\text{LEP}} = 137 \,\text{pb}^{-1}$. While HL-LHC will not be able to improve the LEP bound due to systematic uncertainties~\cite{Cobal:2020hmk}, FCC-ee has an estimated projected sensitivity to branching ratios of $2\times10^{-11}$~\cite{Cobal:2020hmk}, for an integrated luminosity $L_{\text{FCC-ee}} = 150 \,\text{ab}^{-1}$. 

Such projection can also be understood via a naive luminosity scaling. Since the main SM background $Z\to \gamma \nu\bar{\nu}$ has a branching ratio of $7.2\times 10^{-10}$ \cite{Denizli:2025pzf}, and considering that, for a background-free process, the sensitivity scales as $1/L$, whereas in the presence of background it scales as $1/\sqrt{L}$, a naive estimate gives
\be
\B_{\text{FCC-ee}}(Z\to \gamma X) \approx \B_{\text{SM}}(Z\to \gamma \nu\bar{\nu})\sqrt{\frac{L_{\text{LEP}}}{L_{\text{FCC-ee}}}\frac{\B_{\text{LEP}}(Z\to \gamma X)}{\B_{\text{SM}}(Z\to \gamma \nu\bar{\nu})}} = 2.5\times 10^{-11} \, ,
\ee
in good agreement with the detailed study in Ref.~\cite{Cobal:2020hmk}. 

\subsubsection{$K \to \pi E_{\rm miss}$}

Recently, NA62 has reached the first observation of the $K^+\to\pi^+\nu\bar\nu$ decay with a significance above $5\sigma$ with data from 2016--2024~\cite{NA62:2024pjp,LaThuile_2026_update}.
The measured central value $\BR(K^+\to\pi^+\nu\bar\nu)=9.6\times10^{-11}$ is well compatible to the SM prediction~\cite{Buras:2022wpw,DAmbrosio:2022kvb,Anzivino:2023bhp} within the $1\sigma$ uncertainty.
The 2016--2022 dataset has been interpreted by the collaboration in terms of a search for $K^{+}\to\pi^{+}X$~\cite{NA62:2025upx}, where $X$ is an invisible state, in the mass range $m_X\in[0,110]\cup[150,260]$ MeV.
In the intermediate range $m_X\in[110,150]$ MeV, the background is enhanced by the $K^+\to\pi^+\pi^0$ decay and therefore bounds on $\BR(K^{+}\to\pi^{+}X)$ come from a recast of $\pi^0\to E_{\rm miss}$ searches~\cite{NA62:2020pwi,NA62:2025upx}.
For masses larger than $260$ MeV, where the experimental search is affected by a large background due to the 3-pion decay of the charged kaon, no bounds are available.
A recast of the full 2016--2024 dataset has been performed by Ref.~\cite{Guadagnoli:2025xnt}.

The KOTO experiment at J-PARC instead targets the rare decay $K_{L}\rightarrow\pi^{0}\nu\bar{\nu}$~\cite{KOTO:2024zbl}.
Zero events were observed in the signal region and upper limits were established on $\BR(K_{L}\to\pi^{0}\nu\bar{\nu})<2.2\times10^{-9}$ at $90\%$ C.L., to be compared with a SM value around $3\times10^{-11}$~\cite{Buras:2022wpw,DAmbrosio:2022kvb,Anzivino:2023bhp}, as well as on $\BR(K_{L}\to\pi^{0}X)$ for a set of mass hypotheses $m_{X}$, where $X$ is an invisible state in the mass range $m_X\in[0,260]$ MeV.

\begin{figure*}[t!]
  \centering
  \includegraphics[width=1.0\textwidth]{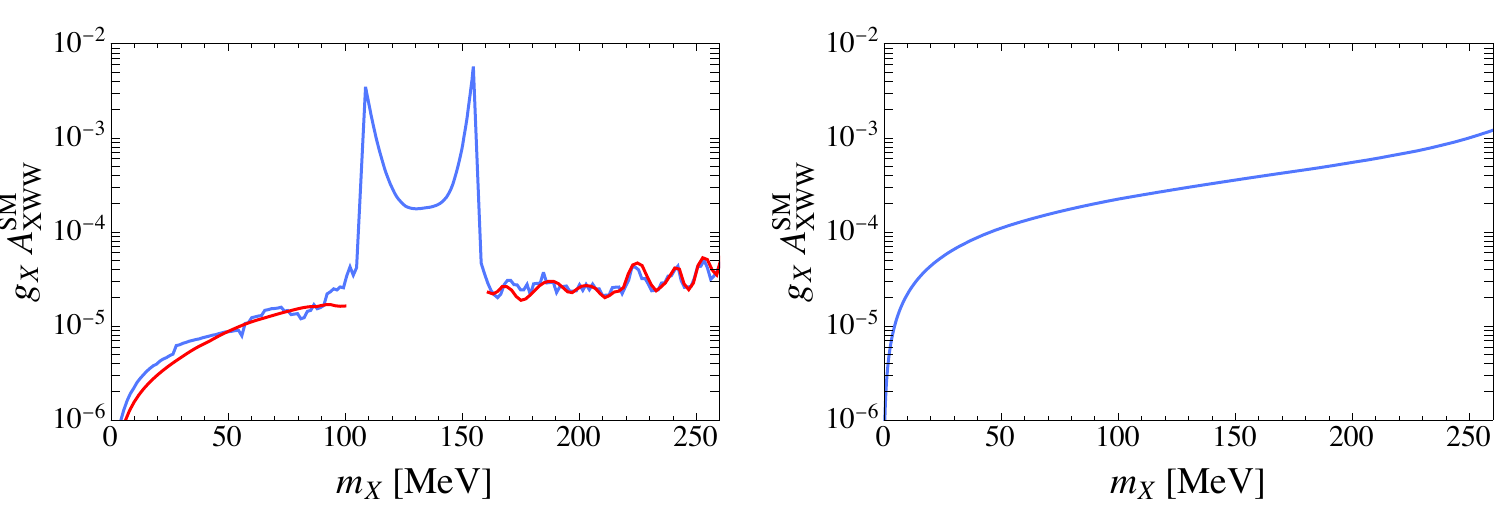}
  \caption{
  Upper limits (blue lines) at $90\%$ C.L.\ on $g_X{\cal A}_{XWW}^{\rm SM}$ from the NA62 search~\cite{NA62:2025upx} (left, based on the 2016--2022 dataset) and the KOTO search~\cite{KOTO:2024zbl} (right).
  The red line in the left panel is the upper limit from the recast of the full 2016--2024 dataset~\cite{NA62:2025upx} of NA62 from Ref.~\cite{Guadagnoli:2025xnt}.
  }
  \label{fig:Ktopi}
\end{figure*}

Within our framework, the above bounds on the $K\to\pi X$ decays can be turned into an upper limit on $g_X {\cal A}_{XWW}^\text{SM}$ by means of Eq.~\eqref{eq:gxididj}, as shown in Fig.~\ref{fig:Ktopi}.
In the particular case of $\tau$-flavor gauge symmetry, the resulting limits from NA62 (with the 2016--2022 dataset) and KOTO are shown, 
respectively in light pink and light blue,  
in Fig.~\ref{fig:pospelov}.

\subsubsection{$B \to \pi E_{\rm miss},\rho E_{\rm miss}$}

The Belle collaboration obtained upper limits at $90\%$ C.L.~on $b\to d\nu\bar\nu$ decays~\cite{Belle:2017oht}, which remain the most stringent ones reported to date.
In this case, the full event kinematics is not reconstructed, and thus the $q^{2}$ dependence is not available. 
Therefore, only global upper limits exist to constrain models with missing energy signature.
Collectively, the bounds at 90\% C.L.~read as
\begin{align}
\label{eq:bellelimit1}
\mathcal{B}(B^{+}\to\rho^{+}E_{\rm miss}) &< 3.0\times10^{-5} \ , \\
\label{eq:bellelimit2}
\mathcal{B}(B^{+}\to\pi^{+}E_{\rm miss}) &< 1.5\times10^{-5} \ , \\
\label{eq:bellelimit3}
\mathcal{B}(B^{0}\to\rho^{0}E_{\rm miss}) &< 4.0\times10^{-5} \ , \\
\label{eq:bellelimit4}
\mathcal{B}(B^{0}\to\pi^{0}E_{\rm miss}) &< 9.0\times10^{-4} \ ,
\end{align}
to be compared to the SM predictions of the neutrino decay modes~\cite{Bause:2021cna}
\begin{align}
\BR(B^+\to \pi^+ \nu\bar\nu) &\simeq 1.2\times10^{-7} \ , \\
\BR(B^0\to \pi^0 \nu\bar\nu) &\simeq 5.6\times10^{-8} \ , \\
\BR(B^+\to \rho^+ \nu\bar\nu) &\simeq 4.9\times10^{-7} \ , \\
\BR(B^0\to \rho^0 \nu\bar\nu) &\simeq 2.3\times10^{-7} \ .
\end{align}
The resulting limits are represented by the blue line   
in Fig.~\ref{fig:pospelov}.
Future investigations of these decay modes may be possible at Belle II. 

\subsubsection{$D \to \pi E_{\rm miss}$}

Searches for $D^+\to(\tau^+\to\pi^+\bar\nu)\nu$ at CLEO~\cite{CLEO:2008ffk} were recast in Ref.~\cite{MartinCamalich:2020dfe} as a constraint on a light ($m_a\simeq 0$) axion coupling mediating $c\to u$ transitions. We reinterpret that result in our setup by rescaling it to 
(the longitudinal component of)
a light vector $X$, which is valid for $m_X$ below the experimental mass resolution, $m_X\lesssim 100~\mathrm{MeV}$~\cite{CLEO:2008ffk}. 
The resulting limit, represented by the orange line   
in Fig.~\ref{fig:pospelov}, is rather weak when compared  
to those stemming from $d_i \to d_j E_{\rm miss}$ transitions, 
mostly due to the relative 
$(m_b/m_t)^4$ suppression in the decay rate (cf.~\eqs{eq:gxididj}{eq:gxiuiuj}). 

Sensitivity projections for analogous missing-energy searches at BESIII \cite{BESIII:2019vhn,BESIII:2024vlt} and the 
Super Tau-Charm Facility (STCF) \cite{Ai:2025xop} can be found in Refs.~\cite{MartinCamalich:2020dfe,MartinCamalich:2025srw}.

\subsubsection{$B \to K^{(*)} E_{\rm miss}$}

\begin{figure*}[t!]
  \centering
  \includegraphics[width=1.00\textwidth]{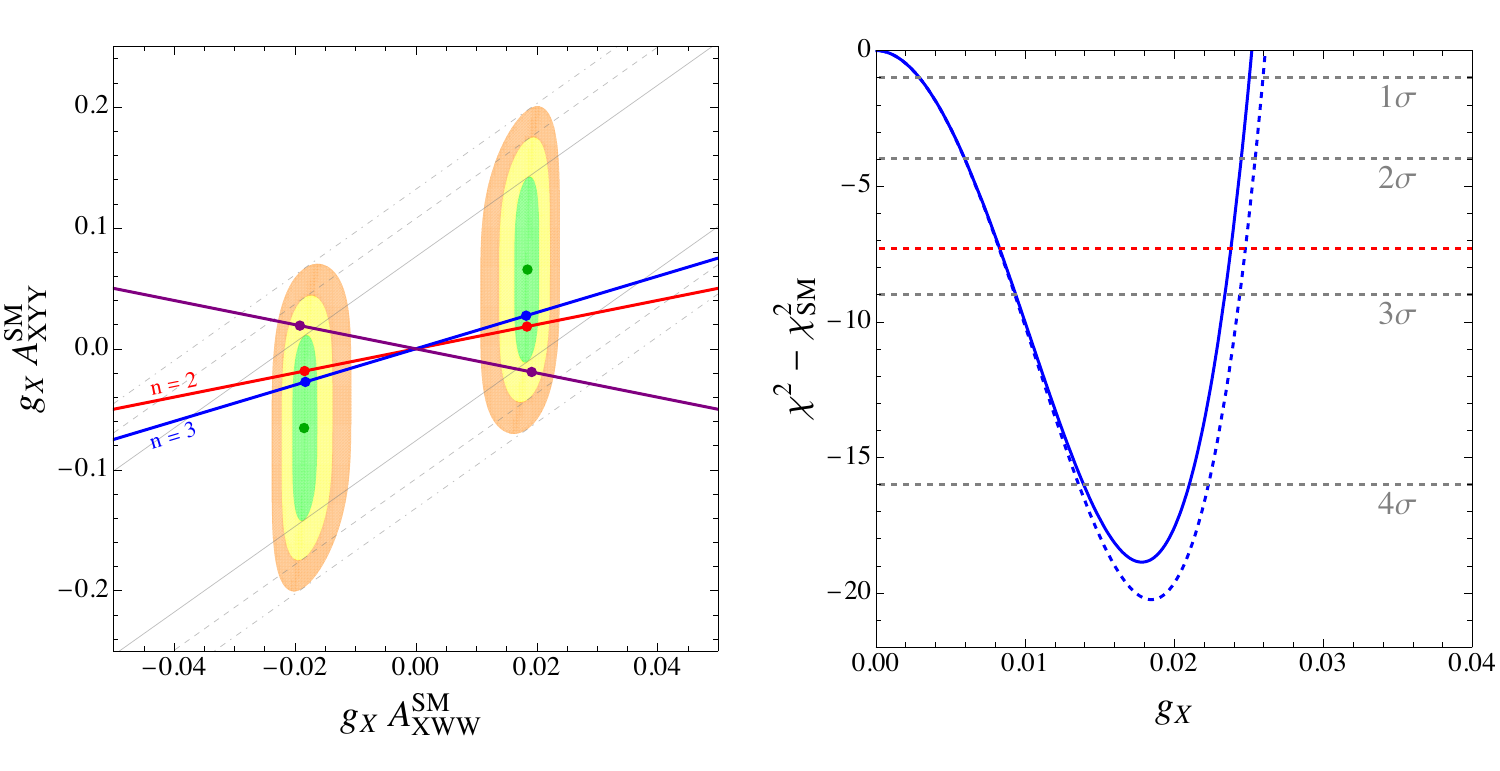}
  \caption{\emph{Left panel:} Regions of constant $\Delta\chi^2\equiv\chi^2-\chi^2_{\rm min}$ in the
$(g_X{\cal A}_{XWW}^{\rm SM},\,g_X{\cal A}_{XYY}^{\rm SM})$ plane. The green, yellow, and orange bands correspond to the
$1\sigma$, $2\sigma$, and $3\sigma$ regions obtained from the combined fit to $B\to K^{(*)}X$ and
$Z\to\gamma X$. For comparison, the solid, dashed, and dot-dashed gray contours show the corresponding $1\sigma$, $2\sigma$, and $3\sigma$
regions from $Z\to\gamma X$ alone. The red and blue curves illustrate the predictions of the $L$-only and $L=-4R$ models of
Ref.~\cite{DiLuzio:2025qkc}, while the purple curve corresponds to the vectorial model from gauged $\tau$-flavor symmetry. The colored dots mark the best-fit points of each model.
\emph{Right panel:} Improvement with respect to the SM as a function of $g_X$ for the gauged $\tau$-flavor symmetry model 
with $m_X = 2.1$ GeV. The solid curve includes
$B\to K^{(*)}X$ and $Z\to\gamma X$, while the dashed curve omits $Z\to\gamma X$. The gray dashed horizontal lines indicate the
$1\sigma$, $2\sigma$, and $3\sigma$ thresholds for a $\chi^2$ distribution with one degree of freedom. The red dashed line shows, for comparison,
the significance over the SM for an interpretation of the Belle~II excess in terms of neutrino emission.}
  \label{fig:plot}
\end{figure*}

Interestingly, the Belle~II experiment has found an excess in the recent $B^+ \to K^+ E_\text{miss}$ measurement~\cite{Belle-II:2023esi}. Assuming that the missing-energy signature is only due to neutrinos, the excess has a significance of $2.7\sigma$ over the SM prediction~\cite{Parrott:2022zte,Becirevic:2023aov}. 
However, Refs.~\cite{Altmannshofer:2023hkn,Bolton:2024egx,Bolton:2025fsq} have shown that the Belle~II excess could be potentially fit by a two-body decay involving an invisible state with mass $(2.1\pm0.1)$ GeV, 
while being compatible with the absence of evidence for $B \to K^{*} E_\text{miss}$ 
at BaBar~\cite{BaBar:2013npw}, with a significance of $4.5\sigma$ over the SM.\footnote{For a more conservative approach, Refs.~\cite{Abumusabh:2025zsr,Gartner:2026clx} recast the Belle~II data as upper limits on $\BR(B^+\to K^+X)$, shown by the green region in \fig{fig:pospelov}.}

Matching the WZ-induced couplings in \eq{eq:gxididj} onto the parametrization of Refs.~\cite{Bolton:2024egx,Bolton:2025fsq},
\be
\label{eq:defgSgP}
g_S\equiv-\frac{m_b-m_s}{2}\frac{g_{\xi sb}}{m_X} \ , \ \ \ 
g_P\equiv\frac{m_b+m_s}{2}\frac{g_{\xi sb}}{m_X} \,,
\ee
the global analysis of the $b\to s E_{\rm miss}$ data performed in Refs.~\cite{Bolton:2024egx,Bolton:2025fsq} yields the best-fit values
\begin{align}
\label{eq:bfgS}
\left|g_S\right|&=(1.6\pm0.2)\times10^{-8} \, , \\
\label{eq:bfgP}
\left|g_P\right|&<2.5\times10^{-8} \text{ (at 90\% C.L.)} \, .
\end{align}
Note that the expressions in \eq{eq:gxididj} are obtained in the regime where the equivalence theorem applies. Hence, 
we expect corrections of order $m_X^2/m_B^2 \approx 15\%$ 
at the best-fit point.

Expressing the constraints on $B\to K^{(*)}X$ and $Z\to \gamma X$ in terms of the SM anomaly traces ${\cal A}_{XYY}^\text{SM}$ and ${\cal A}_{XWW}^\text{SM}$, we construct a $\chi^2$ variable for these observables.\footnote{The interaction in Eq.~\eqref{eq:xidZdAlag} also induces $B\to\pi X$ and $B\to\rho X$ decays, for which only upper limits are available from Belle~\cite{Belle:2017oht}. We include these constraints and find that they have a negligible impact on our results. Moreover, for the benchmark values of the gauge coupling, on-shell $X$ emission is kinematically forbidden in $K\to\pi E_{\rm miss}$ decays; we have checked that the corresponding off-shell contribution is negligible.}
In particular, we take Gaussian terms for the \emph{squared} couplings in Eqs.~\eqref{eq:defgSgP}--\eqref{eq:bfgS}, since the Belle~II signal rate scales with their squares.

The left panel of \fig{fig:plot} shows the $1\sigma$, $2\sigma$, and $3\sigma$ regions in the $(g_X{\cal A}_{XWW}^{\rm SM},\,g_X{\cal A}_{XYY}^{\rm SM})$ plane. Note that for a fixed choice of gauge symmetry, the SM anomalous traces are fixed, so that $g_X$ is the only free parameter in the fit. The right panel displays the improvement with respect to the SM as a function of $g_X$ for the gauging of the accidental SM $\tau$-lepton flavor symmetry.
The best-fit point for the $\tau$-flavor model (purple dot in the left panel) corresponds to $g_X\simeq 0.018$. It lies slightly outside the $1\sigma$ region in the $(g_X{\cal A}_{XWW}^{\rm SM},\,g_X{\cal A}_{XYY}^{\rm SM})$ plane because the predicted rate $\BR(Z\to\gamma X)=5.6\times10^{-7}$ is close to the current experimental limit. Nevertheless, the model yields a substantial improvement over the SM, reaching a $4.3\sigma$ preference, as shown in the right panel of \fig{fig:plot}.

More recently, Belle has published bounds on $B^+\to K^+ X$, with $X$ identified as missing energy~\cite{Belle-II:2026tyb}. However, mass regions around $\pi^0$, $K^{0(*)}$, $D^{0(*)}$, $\eta_1$, $\chi_{c1}$ and $\psi(2S)$ masses, which include the region around 2 GeV, are not constrained due to the veto of the Belle analysis.

On a final note, future measurements of the $B\to K^{*}X$ would be able to discriminate among different explanations of the Belle II, as discussed e.g.~in Ref.~\cite{Bolton:2025fsq}. The left-handed structure in Eq.~\eqref{eq:gxididj} implies a definite prediction for the vector model: at the Belle~II best-fit point one finds $\BR(B^+\to K^{*+}X)=1.1\times10^{-5}$, which can be tested in future Belle~II searches.

\subsection{IR--UV interplay}

\begin{figure*}[t!]
  \centering
  \includegraphics[width=1.0\textwidth]{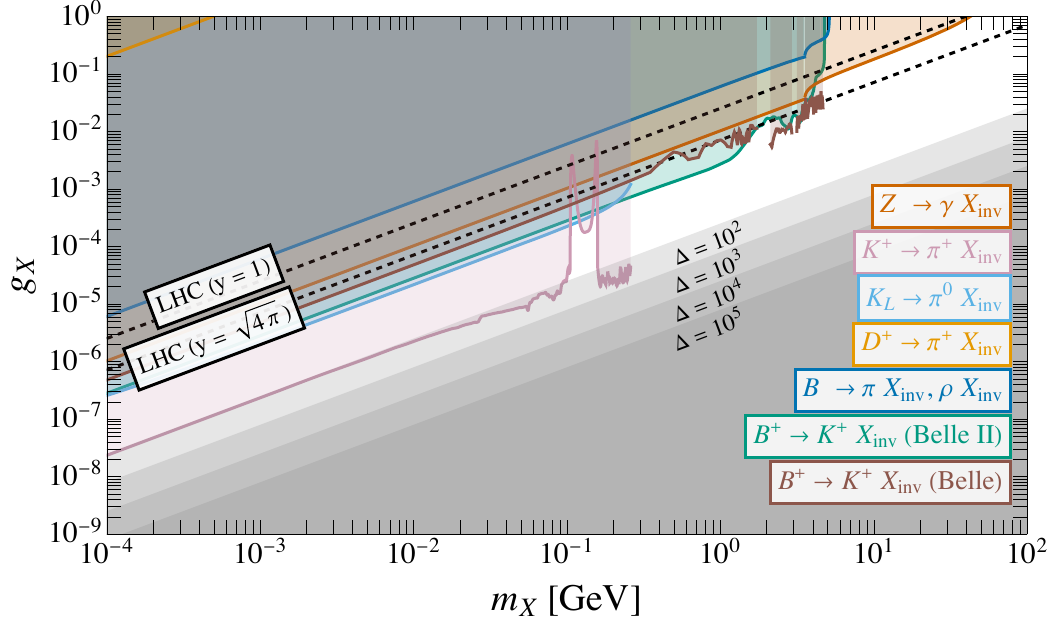}
  \vspace{0.5cm}
  \includegraphics[width=1.0\textwidth]{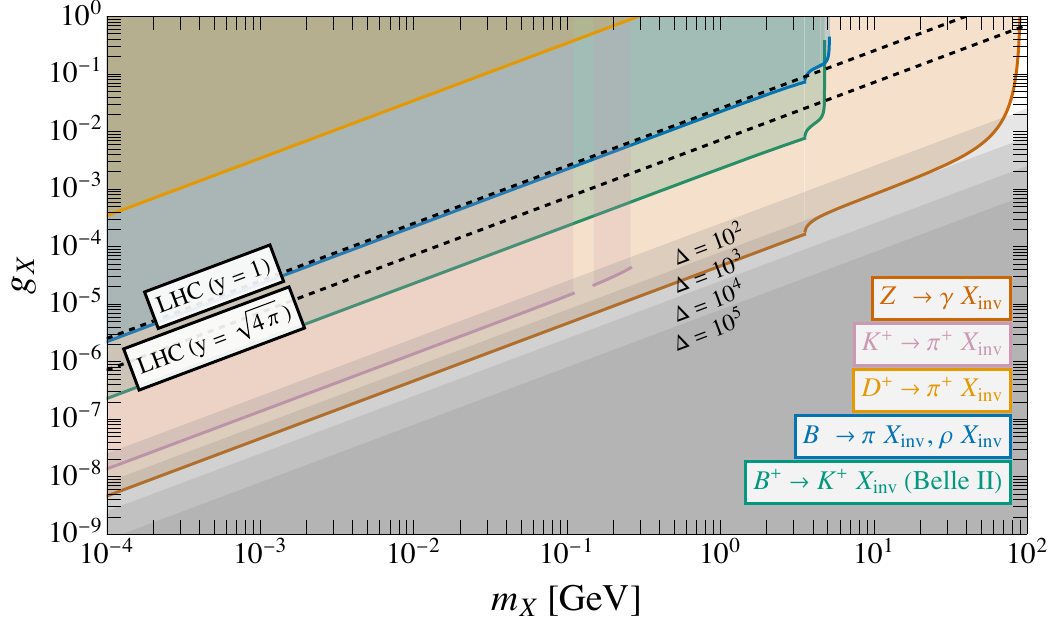}
  \caption{\emph{Upper panel:} Lower limits on the gauge coupling $g_X$ at $90\%$ C.L.\ for the gauged $\tau$-flavor symmetry benchmark, derived from missing-energy searches in $Z\to\gamma X$ at LEP (L3)~\cite{L3:1997exg} (dark orange), $K^+\to\pi^+X$ at NA62~\cite{NA62:2025upx} (light pink), $K_L\to\pi^0X$ at KOTO~\cite{KOTO:2024zbl} (light blue), $D\to\pi X$ at CLEO~\cite{CLEO:2008ffk} (orange), 
  $B\to\pi X,\rho X$ at Belle~\cite{Belle:2013tnz} (blue), 
  and 
  $B^+\to K^+X$ at Belle~\cite{Belle-II:2026tyb} (brown) and at Belle~II~\cite{Belle-II:2023esi} (bluish green, recast in Ref.~\cite{Abumusabh:2025zsr}). 
  Also shown are indicative bounds from direct LHC searches (black) and finite-naturalness estimates (gray bands). 
  \emph{Lower panel:} same as upper panel, considering future projections (dashed lines) for rare meson decays at flavor factories and invisible $Z$ decays at FCC-ee (see text for details). 
  }
  \label{fig:pospelov}
\end{figure*}

Fig.~\ref{fig:pospelov} (upper panel) collects the $90\%$ C.L.\ lower limits on $g_X$ for the gauged $\tau$-flavor symmetry benchmark (${\cal A}_{XYY}^{\rm SM}=-{\cal A}_{XWW}^{\rm SM}=3\alpha_{B+L}=1$), obtained from $Z\to\gamma E_{\rm miss}$ at LEP, kaon decays at NA62 and KOTO, $D$-meson decays at CLEO, and $B$-meson decays at BaBar/Belle/Belle~II. The $B^+\to K^+X$ bound (green, recast from Belle~II) weakens around $m_X\simeq2~\mathrm{GeV}$, mirroring the signal-like preference in the same dataset. In this benchmark, invisible decays (and hence missing-energy signatures) are obtained for $m_X<2m_\tau\simeq3.55~\mathrm{GeV}$.
Assuming, in a minimal setup, that at low energies $X$ couples only to the $\ell^\tau_L$ and $e^\tau_R$ fields,\footnote{The presence of a right-handed $\tau$ neutrino with the same $\tau$-flavor charge, or additional light dark-sector states, could modify the invisible branching ratio of $X$.} the branching ratio into invisible final states is explicitly given by
\be
\BR(X\to E_\text{miss})=\begin{cases}
1 \quad &\text{if $m_X<2m_\tau$\,,} \\
\left[1+2\sqrt{1-\frac{4m_\tau^2}{m_X^2}}\left(1+\frac{2m_\tau^2}{m_X^2}\right)\right]^{-1} \quad &\text{if $m_X\geq2m_\tau$\,.}
\end{cases}
\ee
For comparison, the same figure also shows indicative constraints from direct anomalon searches at the LHC (discussed in Section~\ref{sec:UVmodel} for representative spectra). We translate these into the $(m_X,g_X)$ plane using the parametric relation $m_\Psi = y_\Psi v_X / \sqrt{2} = y_\Psi\, m_X/(\sqrt{2}g_X)$, and impose conservatively $m_\Psi>100~\mathrm{GeV}$, respectively for $y_\Psi = 1$ and $y_\Psi = \sqrt{4\pi}$ (black, dashed lines). With this choice, the indirect bounds from $K\to\pi X$ and $B\to K X$ already dominate over current direct limits.

We also overlay, in gray bands, finite-naturalness estimates from Section~\ref{subsec:finitenaturalness}. Taking $y_\Psi=\sqrt{4\pi}$ as a perturbative upper benchmark and requiring $m_\Psi\lesssim \sqrt{\Delta}\,\mathrm{TeV}$ (corresponding to a tuning $1/\Delta$), one sees that the NA62 constraint from $K^+\to\pi^+X$ is already sensitive to UV completions at the percent-level tuning threshold.

Finally, the lower panel of Fig.~\ref{fig:pospelov} includes projected sensitivities (dashed lines, same color code) for $Z\to\gamma E_{\rm miss}$ at FCC-ee (cf.~\sect{sec:Zdecay}) and for rare meson decays, the latter taken from Ref.~\cite{MartinCamalich:2025srw}.
In the absence of a dedicated study for a massive $X$ for all channels, we overlay these projections using the massless approximation. Consequently, the sensitivity is expected to degrade near kinematic endpoints, due to experimental efficiencies and increasing SM backgrounds. In the case of NA62, we restrict the prospect in the same signal regions of the present analysis based on $K^+\to\pi^+\nu\bar\nu$, outside of which the background dominates over the signal.
Although approximate, these projections provide a useful guide for future rescalings. In particular, the anticipated $2$--$3$ orders-of-magnitude improvement in $Z\to\gamma E_{\rm miss}$ at FCC-ee would probe this framework down to the per-mille tuning level, and is expected to dominate over rare meson decays and direct anomalon searches.

\section{UV models}
\label{sec:UVmodel}

We now present explicit UV completions based on the anomaly-canceling fermion spectra identified in Section~\ref{sec:minanomspect}, using as a motivated benchmark the interpretation of the Belle~II excess in terms of an $X$ boson with mass $m_X=2.1\,\mathrm{GeV}$.

In each model the scalar sector is composed by the SM Higgs boson $H$, uncharged under the new gauge symmetry, and the new complex scalar $\S$, singlet under the SM gauge group and responsable for the SSB of $\U(1)_X$.
By adding a proper term in the scalar potential, $\Delta V (H, \S)$, 
the following VEV configurations 
are generated
\beq 
\vev{H} = \frac{1}{\sqrt{2}} 
\begin{pmatrix}
0 \\ v
\end{pmatrix} \, , \qquad 
\vev{\S} = \frac{v_X}{\sqrt{2}} \, , 
\eeq
with $v \approx 246$ GeV and $v_X$ being the order parameter of $\U(1)_X$ breaking. 
The latter is responsible for the mass of the $\U(1)_X$ gauge boson, $X^{\mu}$, that is
\beq 
\label{eq:massX}
m_{X} = \X_\S g_X v_X \, ,
\eeq
with the covariant derivative defined as 
$D^\mu \S \equiv (\partial^\mu + i g_X \X_\S X^{\mu}) \S$.

Matching the coupling and mass of the gauge boson to the best-fit value of the Belle~II excess sets the VEV of the $\S$ boson as
\begin{align} 
\label{eq:scaleX}
v_X = 
 \frac{120 \, \text{GeV}}{\X_\S} \( \frac{m_X}{2.1 \, \text{GeV}} \) \( \frac{0.018}{{g_X}} \)
\, ,
\end{align}
which depends only on the $\U(1)_X$ charge of $\S$.

This implies that the SM Higgs $H$ and $\S$ acquire VEVs of comparable size. In principle, they may exhibit significant mixing effects through the renormalizable operator $\lambda_{SH}\,\mathcal{S}^{\dagger} \mathcal{S}\, H^{\dagger} H$. The scalar-sector phenomenology then depends strongly on $\lambda_{SH}$ and on the mass of the radial mode of $\mathcal{S}$, as is generically the case in models with an extra gauged $\U(1)$ symmetry. We do not pursue these effects further here. 

Concerning Higgs decays, loop-induced modes such as $h\to\gamma X$, $h\to ZX$, and $h\to XX$ are also present in our setup. We have checked that, in the parameter region relevant for the Belle~II excess, these channels do not yield competitive constraints. 
On the other hand, if Yukawa couplings between anomalons and the SM Higgs are present, they are constrained by Higgs decays such as $h\to\gamma\gamma$ and $h\to Z\gamma$, as well as by electroweak precision tests. The most recent measurements of the Higgs decay rates, normalized to the SM predictions, give $R_{\gamma\gamma}=1.00\pm0.12$~\cite{ATLAS:2019nkf} and $R_{Z\gamma}=1.3^{+0.6}_{-0.5}$~\cite{ATLAS:2025aip}\footnote{While the first evidence for $h\to Z\gamma$ reported by ATLAS and CMS~\cite{CMS:2023mku} showed a $\sim2\sigma$ deviation from the SM expectation, the updated Run-2 result is compatible with the SM.}, while the latest global fits to the oblique parameters yield~\cite{Gfitter} $S=0.05\pm0.11$ and $T=0.09\pm0.13$. The resulting bounds are discussed below for each model considered.

\subsection{1DL--1ML model}
\label{sec:1DL1MLmodel}

We first consider a UV model with one DL pair and one ML state, corresponding to $N_\Psi = 3$. The simplest realization that satisfies the phenomenological requirement (ii) in Section~\ref{subsec:phenomenology} (i.e.~neutral LP in the anomalon sector) is
shown in the first line of Table~\ref{tab:sol1ML1DL},
and involves a $\SU(2)_L$-doublet DL pair $\A_{L,R}$ and a $\SU(2)_L$-triplet ML state $\B_L$, with the same SM quantum numbers as the Higgsino and the wino, respectively. From Eq.~\eqref{eq:scaleX} the corresponding breaking scale is $v_X \simeq 120 \, \text{GeV}$. 
The Lagrangian of the anomalon Yukawa couplings is then given by
\begin{align}
-\lag^\text{Yuk}_{\Psi\S}-\lag^\text{Yuk}_{\Psi H} =&\; y_{AA}\,\bar\A_L \A_R \S +\frac{1}{2}y_{BB}\,\bar\B_L (\B_L)^c \S^* \nonumber \\
+&\, \frac{y_{AB}}{\sqrt{2}}\,(\bar\A_L \sigma^a H) \B_L^{c,a} +\frac{y_{BA}}{\sqrt{2}}\,(\bar\B_L)^a (H^\dagger \sigma^a\A_R) + \text{h.c.} \label{eq:lag1DL1MLm2}
\end{align}
where $a=1,2,3$ is the isospin index in the adjont representation.
After SSB, the Lagrangian~\eqref{eq:lag1DL1MLm2} gives mass to the anomalon fields.
To avoid the stringent bounds on chiral exotic fermions \cite{Barducci:2023zml}, we assume that the anomalon masses arise predominantly from $v_X$, i.e. $m_\A\approx y_{AA}v_\S/\sqrt{2}$ and $m_\B\approx y_{BB}v_\S/\sqrt{2}$.
Requiring perturbative unitarity on the Yukawa couplings~\cite{Allwicher:2021rtd}
\be
y_{AA}\lesssim\sqrt{4\pi} \ , \quad y_{BB}\lesssim \sqrt{\frac{16\pi}{3}} \ ,
\ee
which limits the mass of the ML triplet $m_\B \lesssim 350$ GeV.

In the limit of small SM Higgs mixing, i.e. $y_{AB,BA}\to0$, the mass splitting in the ML triplet is given by radiative corrections, yielding $m_{B^+}-m_{B^0}\approx166$ MeV~\cite{Cirelli:2005uq}. Bounds from disappearing track searches in a wino-like scenario~\cite{CMS:2023mny} then push our model beyond the perturbative limit.
Instead, when contributions from SM Higgs mixing are relevant, the neutral fields $A_L^0,A_R^0,B_L^0$ mix into three Majorana fermions. Direct searches from ATLAS then constrain decays of charginos- and neutralino-like states, pushing the model at the edge of perturbativity~\cite{ATLAS:2024qxh}.

\begin{table}[!ht]
	\centering
	\begin{tabular}{|c|c|c|c|c|c|}
	\hline
	Field & Lorentz & $\SU(3)_C$ & $\SU(2)_L$ & $\U(1)_Y$ & $\U(1)_{X}$ \\ 
	\hline
	$\A_L$ & $(\tfrac{1}{2},0)$ & 1 & 3 & $1$ & $1/12$ \\
	$\A_R$ & $(0,\tfrac{1}{2})$ & 1 & 3 & $1$ & $-1/12$ \\
	$\B_L$ & $(\tfrac{1}{2},0)$ & 1 & 5 & $0$ & $-1/12$ \\
	\hline
	$\N_R^{1}$ & $(0,\tfrac{1}{2})$ & 1 & 1 & $0$ & $1/12$ \\
    $\N_R^{2}$ & $(0,\tfrac{1}{2})$ & 1 & 1 & $0$ & $1$ \\
    \hline
    $\S$ & $(0,0)$ & 1 & 1 & 0 & $1/6$ \\ 
    \hline 
	\end{tabular}	
	\caption{\label{tab:1DL1ML_1} 
	Field content of the 1DL--1ML model in Eq.~\eqref{eq:sol1DL1ML} with $m=3$ and $\alpha_{B+L} = 1/3$.}
\end{table}

Therefore, we also consider a next-to-minimal model based on the solution in Eq.~\eqref{eq:sol1DL1ML} with $m=3$, whose field content is reported in Table~\ref{tab:1DL1ML_1}. In this case, the SSB scale is equal to $v_X \simeq 720\,\text{GeV}$, which allows the anomalon masses to lie at the TeV scale.

The Yukawa Lagrangian of the anomalons is
\be
\label{eq:lag1DL1MLm3}
-\lag^\text{Yuk}_{\Psi\S} =
y_{AA}\bar\A_L  \S \A_R + y_{BB} \bar\B_L \S^* (\B_L)^c + \text{h.c.} \ ,
\ee
where the isospin indices of each term are implicitly understood to be contracted into a $\SU(2)_L$-singlet through Clebsch-Gordan coefficients.
In this case, gauge invariance forbids SM Higgs mixing among the different multiplet (namely $\lag_{\Psi H}^\text{Yuk} = 0 $), so that the fields $\mathcal{A}$ and $\mathcal{B}$ are mass eigenstates. 
The bound of perturbative unitarity, which reads~\cite{Allwicher:2021rtd}
\be
y_{AA}\lesssim\sqrt{\frac{8\pi}{3}} \ , \quad y_{BB}\lesssim \sqrt{\frac{16\pi}{5}} \ ,
\ee
limits masses of the anomalons below $1.47$ TeV and $1.6$ TeV for the DL triplet and ML quintuplet, respectevely. For reference, the naturalness criterion of 1\% fine tuning imposes anomalon mass scale below $3.4$ TeV. 

Both the $\mathcal{A}$ and $\mathcal{B}$ multiplets contain a lightest neutral component, which is stable due to a residual $Z_2$ symmetry acting only on the anomalon fields. As a consequence, bounds from charged tracks are avoided.
Moreover, since the typical mass scale is $\mathcal{O}(\text{TeV})$, current disappearing-track constraints are evaded~\cite{CMS:2023mny}.

The SM-singlet fields $\N_R^{i=1,2}$ in Table~\ref{tab:1DL1ML_1} are introduced to cancel the $[\U(1)_X]$ and $[\U(1)_X^3]$ anomalies.
The SM singlets take mass through the Lagrangian
\be
\label{eq:LagSMsinglets1DLand1ML}
-\lag^\text{Yuk}_{\N\S}-\lag_{\ell\N} = \frac{1}{2}y_{\N} \overline{\N_R^1} \S (\N_R^1)^c + y_{\nu_\tau} \overline{\ell^\tau_L} \tilde{H} \N_R^2  + \text{h.c.} \ ,
\ee
where $y_{\N}$ and $y_{\tau\N}$ can be set real and positive by a global rephasing of the SM-singlet fields, and with $\tilde{H}=i\sigma_2 H^*$.
After SSB, the spectrum of Eq.~\eqref{eq:LagSMsinglets1DLand1ML} contains the Majorana fermion $\N_M\equiv\N_R^1+(\N_R^1)^c$, with mass $m_{\N_M}=y_\N v_\S/\sqrt{2}$, and the $\tau$-flavor neutrino $\nu_\tau\equiv\N_R^2+\nu^\tau_L$ with Dirac mass
$m_{\nu_\tau}=y_{\nu_\tau} v/\sqrt{2}$.
However, a realistic neutrino spectrum requires introducing at least two copies of $\N_R^2$.

Since the neutral component $\A^0$ is stable, 
due to an accidental $Z_2$ symmetry acting on $\A$, it provides a potential dark matter (DM) candidate. 
Given that $g_X \ll g$, its relic abundance is driven by the freeze-out of EW interactions, with $\Omega h^2 \propto m_{\A^0}^2$.  
Given the viable mass range for a wino-like triplet \cite{Bottaro:2022one}, the fraction of the total DM density accounted for by $\A^0$, i.e.~$n_{\A^0}/n_{\rm DM}$, is estimated to be of order $1\%$.
However, direct detection from DM searches highly constrains DM candidates interacting with the $Z$ boson. The most stringent bounds from Ref.~\cite{LZ:2024zvo} require $n_{\A^0}\lesssim10^{-9}n_\text{DM}$, 
thereby ruling out this model under standard cosmological assumptions.

To avoid such constraints, one can explicitly break the accidental $Z_2$ symmetry by means of a higher-dimensional operator, either to force the decay of $\A^0$, or to separate the Majorana components of $\A^0$ by a mass splitting large enough to kinematically evade direct detection (i.e.~inelastic splitting, see e.g.~\cite{Bottaro:2022one}).

On the other hand, $\B^0$ stability poses no issues because it lacks tree-level $Z$ boson interaction. Indeed, the ML quintuplet corresponds to the only viable minimal DM candidate \cite{Cirelli:2005uq,DelNobile:2015bqo} with an extra $\U(1)_X$ dark charge. Since $g_X \ll g$, the DM abundance is reproduced by setting $m_{\B^0} \approx 15$ TeV \cite{Mitridate:2017izz,Bottaro:2021snn}. As shown in \cite{Mitridate:2017izz}, a stable candidate $\B^0$ with $m_{\B^0} \approx 300$ GeV can account for $\lesssim 3\%$ of DM. $B^0$-nucleon loop-induced interactions are purely electroweak \cite{Bottaro:2022one}, pushing the spin independent cross section below the neutrino floor.

Overall, the $m=3$ realization provides a fully viable benchmark scenario: it satisfies anomaly cancellation, remains perturbative up to the SSB scale, evades current collider bounds, and features a stable neutral state without introducing too much fine-tuning for the Higgs mass parameter (cf.~last column in Table~\ref{tab:sol1ML1DL}). In this sense, it represents the minimal UV-complete model of the 1DL--1ML setup that is both phenomenologically safe and experimentally testable in future high-luminosity collider runs.

\subsection{2DL model}
\label{sec:2DLmodel}

In addition, it is worth considering a UV model with two DL anomalon fields, 
as given by the solution in Eq.~\eqref{eq:sol2DL} with $m=1$ and $\alpha_{B+L} = 1/3$.
The field content is explicitly shown in Table~\ref{tab:2DL_1}, from which Eq.~\eqref{eq:scaleX} yields $v_X=120$ GeV.
We generally leave $\X$ unfixed, with the only restriction that no terms mixing anomalon and SM fermion fields are allowed. A possible choice would be for instance $\X=0$.

\begin{table}[!ht]
	\centering
	\begin{tabular}{|c|c|c|c|c|c|}
	\hline
	Field & Lorentz & $\SU(3)_C$ & $\SU(2)_L$ & $\U(1)_Y$ & $\U(1)_{X}$ \\ 
	\hline
	$\A_L$ & $(\tfrac{1}{2},0)$ & 1 & 1 & $1$ & $\X+1/2$ \\
	$\A_R$ & $(0,\tfrac{1}{2})$ & 1 & 1 & $1$ & $\X-1/2$ \\
	$\B_L$ & $(\tfrac{1}{2},0)$ & 1 & 2 & $1/2$ & $\X-1/2$ \\
	$\B_R$ & $(0,\tfrac{1}{2})$ & 1 & 2 & $1/2$ & $\X+1/2$ \\
	\hline
	$\N_R^{1}$ & $(0,\tfrac{1}{2})$ & 1 & 1 & $0$ & $\X-1/2$ \\
    $\N_R^{2}$ & $(0,\tfrac{1}{2})$ & 1 & 1 & $0$ & $-\X-1/2$ \\
    $\N_R^{3}$ & $(0,\tfrac{1}{2})$ & 1 & 1 & $0$ & $1$ \\
    \hline
    $\S$ & $(0,0)$ & 1 & 1 & 0 & $1$ \\ 
    \hline 
	\end{tabular}	
	\caption{\label{tab:2DL_1} 
	Field content of the model with two DL anomalon fields as given by the solution in Eq.~\eqref{eq:sol2DL} with $m=1$ and $\alpha_{B+L} = 1/3$.} 
\end{table}

The Lagrangian of the anomalon Yukawa couplings is then given by
\be
\label{eq:lag2DL1}
-\lag^\text{Yuk}_{\Psi\S}-\lag^\text{Yuk}_{\Psi H} =
\begin{pmatrix}
\overline{\A_L} & \overline{\B_L}
\end{pmatrix}
\begin{pmatrix}
y_{AA} \S & y_{AB} H^\dagger \\
y_{BA} H & y_{BB} \S^*
\end{pmatrix}
\begin{pmatrix}
\A_R \\
\B_R
\end{pmatrix}
 + \text{h.c.} \ ,
\ee
where three of the four phases of $y_{AA,AB,BA,BB}$ can be removed by rephasing the anomalon fields and are therefore unphysical.
Therefore, we can set $y_{AA,BB}>0$ with no loss of generality.  The criterion of perturbative unitarity implies~\cite{Allwicher:2021rtd}
\be
\label{eq:PU2DL}
y_{AA}\lesssim\sqrt{8\pi} \ , \quad y_{BB}\lesssim \sqrt{4\pi} \ .
\ee
Upper bounds on $y_{AB,BA}$ instead follow from EW precision tests and from measurements of $h\to\gamma\gamma$ and $h\to Z\gamma$. For $y_{AA,BB}\sim\mathcal{O}(1)$, these contributions remain negligible provided $y_{AB,BA}\lesssim\mathcal{O}(10^{-1})$. This conclusion persists even when using projected sensitivities for EW precision observables at the HL-LHC and FCC-ee~\cite{deBlas:2016ojx}.

After SSB, both $\A$ and $\B$ acquire masses, and the mass matrix in Eq.~\eqref{eq:lag2DL1} must be diagonalized. To avoid the stringent bounds on chiral exotic fermions \cite{Barducci:2023zml}, we assume that the anomalon masses arise predominantly from $v_X$, i.e.\ $m_\A\approx y_{AA}v_\S/\sqrt{2}$ and $m_\B\approx y_{BB}v_\S/\sqrt{2}$. Yukawa interactions involving the SM Higgs then provide subleading effects, in particular inducing mass splittings between the components of $\B=(\B^+,\ \B^0)^T$.

For small SM Higgs mixing, namely $y_{AB,BA}\lesssim\mathcal{O}(10^{-3})$, the mass splitting among the $\B$ components is dominated by radiative corrections, yielding $m_{\B^+}-m_{B^0}\approx350$ MeV~\cite{Cirelli:2005uq}. Searches for disappearing tracks of the Higgsino-like decay $\B^+\to\pi^+\B^0$ then leads a lower limit on the mass that reads as $m_{B^0}\gtrsim200$ GeV~\cite{CMS:2023mny}, while the perturbative unitarity bound of Eq.~\eqref{eq:PU2DL} yields $m_{B^0}\lesssim300$ GeV.
For larger values of the SM Higgs mixing, the constraint from direct searches depends on the mass splitting and hence on the explicit values of $y_{AB,BA}$.
As a particular case, the limit in which $y_{AB,BA}$ are real and opposite ensures that the splitting $m_{\B^+}-m_{B^0}$ due to SM Higgs mixing at tree level is always positive and can be as large as few GeVs while still being compatible with SM Higgs observables, thus helping to avoid the charged track bounds.
Future searches for disappearing tracks at FCC~\cite{Han:2018wus} or muon colliders~\cite{Capdevilla:2021fmj} will be able to perturbatevely exclude the 2DL model, but only up to splitting values as large as 500 MeV, due to an inherent physical resolution limit of detection.
Meanwhile, soft tracks analysis at HL-LHC and FCC-ee~\cite{CidVidal:2018eel} will test this decay beyond the perturbative limit for values of the splitting larger than 2 GeV.
Hence, mass splittings in the $0.5$--$2~\mathrm{GeV}$ range are expected to remain essentially unconstrained for the foreseeable future.

In the limit of vanishing SM Higgs mixing, i.e.~$y_{AB,BA}\to0$, the charged fermion $\A$ is stable, with the size of its mass controlled by $v_X\approx120$ GeV.
Hence, bounds from charged tracks~\cite{CMS:2016kce} push its Yukawa coupling $y_{AA}$ well beyond the perturbative regime, namely $y_{AA}\gtrsim6$.
A non-vanishing value of the SM Higgs mixing is then required to allow $\A$ to decay into a component of the multiplet $\B^0$ through $\A\to V \B$, with $V=W,Z$ being on or off-shell, assuming $\Delta m \equiv m_\A-m_\B>0$.

The minimal requirement of a prompt decay to avoid charged tracks inside the detector sets a lower limit on the size of the SM Higgs mixing.
If $\Delta m\geq m_V$, the decays $\A\to V\B$ can be prompt compared to the detector resolution at sufficently small values of the SM Higgs mixing, namely $y_{AB,BA}\gtrsim\mathcal{O}(10^{-6})$, to evade bounds from direct searches on $\A$.
If, instead, $\Delta m < m_V$, then the possible decays are $\A\to \B f_i \bar{f}_j$, where $f_{i,j}$ are SM leptons or quarks. Summing all contributions, the decay can be prompt for $|y_{AB,BA}| \gtrsim \mathcal{O}(10^{-5})$.

Regarding the bounds from direct searches for prompt $\A^+\to W^+\B^0,Z\B^+$ decays, a full recast of ATLAS and CMS constraints would be necessary, which is beyond the scope of this work. We notice, however, that present bounds from wino searches are less constraining than disappearing track searches~\cite{ATLAS:2024qxh}. 
Future projections for such searches show that HL-LHC will be able to exclude most of the parameter space for masses $\lesssim\mathcal{O}(500\text{ GeV})$~\cite{CidVidal:2018eel}, with the exception for the region where the mass spliting is a few GeV above the $W$ mass.

The SM-singlet fields $\N_R^{i=1,2,3}$ in Table~\ref{tab:2DL_1} are introduced to cancel the $[\U(1)_X]$ and $[\U(1)_X^3]$ anomalies.
The SM-singlets $\N_R^{i=1,2}$ form a DL pair, while $\N_R^{3}$ acts as the right-handed neutrino for the $\tau$ lepton sector.
The SM-singlets then take mass through the Lagrangian
\be
\label{eq:LagSMsinglets2DL}
-\lag^\text{Yuk}_{\N\S} = y_{\N} \overline{\N_R^1} \S^* (\N_R^2)^c + y_{\nu_\tau} \overline{\ell^\tau_L} \tilde{H} \N_R^3  + \text{h.c.} \ ,
\ee
where $y_{\N}$ and $y_{\tau\N}$ can be set as real and positive by a global rephasing of the SM-singlets, and with $\tilde{H}=i\sigma_2 H^*$.
After SSB, the spectrum of Eq.~\eqref{eq:LagSMsinglets2DL} contains the Dirac fermion\footnote{In the case $\chi=0$, the additional Lagrangian term $-\Delta\lag^\text{Yuk}_{\N\S} = y_{\N_1} \overline{\N_R^1} \S^* (\N_R^1)^c+y_{\N_2} \overline{\N_R^2} \S^* (\N_R^2)^c + \text{h.c.}$ is allowed and $\N_R^{i=1,2}$ mix instead to form two Majorana fermions after SSB.} $\N_D\equiv\N_R^1+(\N_R^2)^c$, with mass $m_{\N_D}=y_\N v_\S/\sqrt{2}$, and the $\tau$-flavor neutrino $\nu_\tau\equiv\N_R^3+\nu^\tau_L$ with Dirac mass
$m_{\nu_\tau}=y_{\nu_\tau} v/\sqrt{2}$.
A realistic neutrino spectrum requires however the introduction of at least two copies of $\N_R^3$.

Given the charge assignment in Table~\ref{tab:2DL_1}, 
the following operator is also allowed 
\be
-\lag^\text{Yuk}_{\Psi\N H} =y_\text{mix}\bar{\B}_L H \N_R^1 + \tilde{y}_\text{mix}\bar{\B}_R H (\N_R^2)^c + \text{h.c.} \ ,
\ee
which breaks the accidental $Z_2$ symmetry of the Lagrangian~\eqref{eq:lag2DL1}, acting on the anomalon fields.
In the limit $y_\text{mix},\tilde{y}_\text{mix}\to 0$, the $Z_2$ symmetry is restored, rendering the anomalon LP (identified above with the neutral component $\B^0$) stable and thus leaving an irreducible relic abundance. Following Sec.~\ref{sec:1DL1MLmodel}, the corresponding DM fraction is estimated to be $n_{\B^0}/n_\text{DM}\sim(1$--$10)\%$, whereas direct-detection limits~\cite{LZ:2024zvo} require $n_{\B^0}/n_\text{DM}\lesssim 10^{-9}$. Non-vanishing values of $y_\text{mix},\tilde{y}_\text{mix}$ are therefore needed to break the accidental $Z_2$ symmetry and allow $\B^0$ to decay. Requiring the decay to occur before Big Bang nucleosynthesis, conservatively $\tau_{\B^0}\lesssim 1\,\mathrm{s}$, implies $y_\text{mix},\tilde{y}_\text{mix}\gtrsim 10^{-12}$. For collider searches and anomalon masses, the impact of $y_\text{mix},\tilde{y}_\text{mix}$ remains negligible up to $\mathcal{O}(10^{-4})$, so that the discussion above is unchanged.

\section{Conclusion}
\label{sec:conclusion}

We have studied light spin-1 gauge bosons, $X$, coupled to EW-anomalous SM currents and developed a general phenomenological framework for their missing-energy signatures. In this setup $X$ decays predominantly into neutrinos, for instance in models based on a $\tau$-flavor symmetry whenever $m_X<2m_\tau$, so that the leading probes are rare processes with invisible final states.
For generic charge assignments, anomaly cancellation requires new fermions that are chiral under $\U(1)_X$ and carry EW quantum numbers. When their masses predominantly arise from $\U(1)_X$ breaking, integrating them out yields an infrared EFT supplemented by Wess--Zumino operators whose coefficients are fixed by mixed-anomaly matching. 

Within this EFT, WZ operators induce $Z\to\gamma X$ at tree level, while the $XW\partial W$ operator generates loop-induced flavor-changing couplings following a predictive minimal-flavor-violation pattern. We have surveyed the resulting missing-energy probes in radiative $Z$ decays and in rare $K$, $D$, and $B$ transitions, and summarized the corresponding bounds in the $(m_X,g_X)$ plane for representative benchmarks. In particular, we emphasized the complementarity between $Z\to\gamma E_{\rm miss}$ and rare-meson searches, and illustrated the framework in the gauged $\tau$-flavor benchmark relevant to Belle~II and NA62/KOTO. 

On the UV side, we classified minimal electroweak-anomalon spectra solving the mixed-anomaly conditions for $N_\Psi\leq 4$, and combined phenomenological requirements with finite-naturalness considerations to constrain the anomalon mass scale. In the parameter region motivated by missing-energy searches, these considerations typically point to anomalon masses in the sub-TeV to few-TeV range, making direct searches an essential part of the overall picture. This interplay is summarized in Fig.~\ref{fig:pospelov}, where we also overlay indicative direct-search constraints in the $(m_X,g_X)$ plane, 
as well as finite-naturalness estimates. 

Finally, we included projected sensitivities for $Z\to\gamma E_{\rm miss}$ 
and for rare meson decays, highlighting the potential of future data to sharpen the IR--UV connection in this framework.

\section*{Acknowledgments}

LDL and MN are supported
by the 
European Union -- Next Generation EU and
by the Italian Ministry of University and Research (MUR) 
via the PRIN 2022 project n.~2022K4B58X -- AxionOrigins.
The work of CT has received funding from the French ANR, under contracts ANR-19-CE31-0016 (`GammaRare') and ANR-23-CE31-0018 (`InvISYble'), that he gratefully acknowledges.

\appendix

\section{Classification of anomalon setups} \label{App:anomalonsetups}

In this Appendix we classify solutions to the anomaly-cancellation conditions in Eqs.~\eqref{eq:anom3}--\eqref{eq:anom5} for $N_\Psi\leq 4$. We proceed by constructing spectra as combinations of ML fields and DL pairs, in accordance with the phenomenological criterion (i). Once solutions satisfying the anomaly constraints are obtained, we then identify the subsets that also fulfill criteria (ii) and (iii). We finally discuss the generaliztion of this classification to colored anomalon fields.

\subsection*{$N_\Psi=1$}

This case is excluded as Eq.~\eqref{eq:anom4} needs at least a DL pair to be solved.

\subsection*{$N_\Psi=2$} \label{Npsi = 2}

In this case we have one DL pair, hence $n_1 = n_2$ and $\Y_1 =  -\Y_2$.
Eq.~\eqref{eq:anom5} requires $\X_1=\X_2$ or $\X_1=-\X_2$, with the latter to be discarded otherwise $\Psi_1$ and $\Psi_2$ are vector-like under the full gauge group.
However, Eq.~\eqref{eq:anom4} then yields $\X_{2} > 0$, while instead Eq.~\eqref{eq:anom3} requires $\X_2 < 0$ to be solved. Given this inconsistency, we conclude that no solution exists in this case. 

\subsection*{$N_\Psi=3$}
\label{subApp:N=3}

In this case, we have a ML field and a DL pair, i.e.
\be
\Psi_1\sim\left(n, 0,\X_1\right) \ , \quad
\Psi_2\sim\left(m, \Y,\X_2\right) \ , \quad
\Psi_3\sim\left(m, -\Y,\X_3\right) \ ,
\ee
where we set $\Y>0$ with no loss of generality.
Eq.~\eqref{eq:anom5} requires $\X_2=\X_3$ or $\X_2=-\X_3$, with the latter to be discarded otherwise $\Psi_2$ and $\Psi_3$ are vector-like under the full gauge group. Eqs.~\eqref{eq:anom3}--\eqref{eq:anom4} then lead to
\be
\X_1 = 3\alpha_{B+L}\frac{1-m^2-12\Y^2}{2n(n^2-1)\Y^2} \ , \quad \X_2=\X_3 = \frac{3\alpha_{B+L}}{4m\Y^2} \ .
\ee
In order to give mass to the ML field and to the DL pair with a single $\S$ boson, one requires $\X_2+\X_3=2\X_1$ or $\X_2+\X_3=-2\X_1$. The former condition admits a solution only for $\Y^2<0$, while the latter yields
\be
\label{eq:Y_1ML_1DL}
\Y=\sqrt{\frac{n(n^2-1)-2m(m^2-1)}{24m}} \ .
\ee
The ans\"atze $n-2m=-1,0,1$ produce three sets of infinite solutions with integer or semi-integer hypercharge values, labelled by the integer $m$.
These sets are:
\begin{itemize}
\item $n=2m+1$ with $m\geq1$:
\be
\Psi_1\sim\left(2m+1, 0,-\frac{3\alpha_{B+L}}{m(m+1)^2}\right) \ ,
\quad
\Psi_{2,3}\sim\left(m, \pm\frac{m+1}{2},\frac{3\alpha_{B+L}}{m(m+1)^2}\right) \ .
\ee
Here, no Yukawa term with the SM Higgs or mixing term with SM fermions (or SM-singlets) is allowed for any value of $m$.
The LP of the DL pair, which does not containt a neutral component, is then a stable charged particle, in contrast with the phenomenological requirement (ii).

\item $n=2m$ with $m\geq1$ and even, in order to avoid Witten anomaly:
\be
\Psi_1\sim\left(2m, 0,-\frac{3\alpha_{B+L}}{m^3}\right) \ ,
\quad
\Psi_{2,3}\sim\left(m, \pm\frac{m}{2},\frac{3\alpha_{B+L}}{m^3}\right) \ .
\ee
We discard this solution since the ML field is a pseudoreal reprensation under $\SU(2)_L$, thus its SM-singlet mass term identically vanishes.

\item $n=2m-1$ with $m\geq2$:
\be
\label{eq:sol1DL1ML}
\Psi_1\sim\left(2m-1, 0,-\frac{3\alpha_{B+L}}{m(m-1)^2}\right) \ ,
\quad
\Psi_{2,3}\sim\left(m, \pm\frac{m-1}{2},\frac{3\alpha_{B+L}}{m(m-1)^2}\right) \ .
\ee
Both the ML field and DL pair contain a neutral component for any value of $m$.
For $m>2$ no Yukawa term with the SM Higgs or mixing term with SM fermions (or SM-singlets) is allowed.
Instead, for $m=2$, one can write a Yukawa term with the SM Higgs among the anomalon fields and, depending on the SM-singlets $\U(1)_X$ charges, the DL pair and a SM-singlet.
On other hand, no mixing with SM fermions is allowed for any value of $m$.
\end{itemize}
Outside of the above sets, we check by an explicit scan that there are no solutions with integer electric charges in the range $n,m\leq10$, or that there are no solutions satisfying (ii) in the same range.

\subsection*{$N_\Psi=4$} \label{sec:Npsi4}
\label{subApp:N=4}

In this case, we have two possibilities: two DL pairs, or a DL pair and two ML fields.
First, we consider two DL pairs defined as
\be
\begin{split}
\Psi_1\sim\left(n, \Y_1,\X_1\right) \ ,
&\quad
\Psi_2\sim\left(n, -\Y_1,\X_2\right) \ , \\
\Psi_3\sim\left(m, \Y_3,\X_3\right) \ ,
&\quad
\Psi_4\sim\left(m, -\Y_3,\X_4\right) \ ,
\end{split}
\ee
where $\Y_1>0$ with no loss of generality.
In order to give mass to the DL pairs with a single $\S$ boson, we require $\X_1+\X_2=\X_3+\X_4$ or $\X_1+\X_2=-\X_3-\X_4$. 
With the former condition, Eq.~\eqref{eq:anom3} requires $\X_1+\X_2 < 0$ while Eq.~\eqref{eq:anom4} leads to $\X_1+\X_2 >0$, so no solution is found.
Assuming instead $\X_1+\X_2=-\X_3-\X_4$, Eqs.~\eqref{eq:anom3}--\eqref{eq:anom5}
leads to an infinite set of solutions parametrized by $n$, $m$ and two of the $\U(1)_X$ charges.
Explicity,
\begin{align}
\Y_1&=\sqrt{\frac{m(m-n)(m^2+n^2+nm-1)}{12n(m-p^2n)}} \ , \label{eq:Y1_2DL}\\
\Y_3=\frac{np}{m}\Y_1&=p\sqrt{\frac{n(m-n)(m^2+n^2+nm-1)}{12m(m-p^2n)}} \ , \label{eq:Y3_2DL}\\
\X_1+\X_2=-\X_3-\X_4&=\frac{18\alpha_{B+L}}{m(m^2-1)-n(n^2-1)} \ ,
\end{align}
where we introduced $p=(\X_1-\X_2)/(\X_3-\X_4)$.
Solutions with (semi-)integer or rational values of the hypercharges $\Y_1$ and $\Y_3$ can be systematically searched for by solving a generalized Pell's equation as described in 
\cite{barbeau2003pell,scacco026diophantine}.  
A particular subset of solutions admits integer electric charges and is defined by the relations $n=m+1$ and $\X_1-\X_2=\X_3-\X_4$, i.e.
\be
\label{eq:sol2DL}
\begin{split}
\Psi_1\sim\left(m+1, \frac{m}{2},\X - \frac{3\alpha_{B+L}}{m(m+1)}\right) \, ,
&\quad
\Psi_2\sim\left(m+1, -\frac{m}{2},-\X - \frac{3\alpha_{B+L}}{m(m+1)}\right) \, , \\
\Psi_3\sim\left(m, \frac{m+1}{2},\X + \frac{3\alpha_{B+L}}{m(m+1)}\right) \, ,
&\quad
\Psi_4\sim\left(m, -\frac{m+1}{2},-\X + \frac{3\alpha_{B+L}}{m(m+1)}\right) \, ,
\end{split}
\ee
which is parametrized by $m$ and a common shift of the $\U(1)_X$ charges labelled as $\X$.
The fields $\Psi_{1,2}$ of Eq.~\eqref{eq:sol2DL} contain a neutral component, while $\Psi_{3,4}$ do not. However, a Yukawa term with the SM Higgs connecting the two DL pairs is always allowed for any value of $\X$ and $m$.
We checked by an explicit scan that, outside of the subset in Eq.~\eqref{eq:sol2DL}, no solutions satisfying (ii) are possible in the range $n,m\leq10$.

Next, we consider a DL pair and two ML fields, defined as
\be
\begin{split}
\Psi_1\sim\left(n_1, 0,\X_1\right) \, ,
&\quad
\Psi_2\sim\left(n_2, 0,\X_2\right) \, , \\
\Psi_3\sim\left(m, \Y,\X_3\right) \, ,
&\quad
\Psi_4\sim\left(m, -\Y,\X_4\right) \, ,
\end{split}
\ee
where $\Y>0$ and $n_2 \geq n_1>1$ with no loss of generality.
Eq.~\eqref{eq:anom5} yields again $\X_3 = \X_4$, while the condition of a single $\S$ boson giving mass to the ML fields and to the DL pair requires $|\X_1| = |\X_2| = |\X_3|$.
Depending on the relative sign among the charges, we obtain different scenarios:
\begin{itemize}
\item  $\X_1 = \X_2 = \X_3$: The combination of Eqs.~\eqref{eq:anom3}--\eqref{eq:anom4} leads to the condition $\Y^2<0$, thus no solution occurs;

\item  $\X_1 = -\X_2 = -\X_3$: Eqs.~\eqref{eq:anom3}--\eqref{eq:anom4} lead to
\be
\Y=\sqrt{\frac{n_1(n_1^2-1)-n_2(n_2^2-1)-2m(m^2-1)}{24m}} \, ,
\quad
\X_3 = \frac{3\alpha_{B+L}}{4m\Y^2}>0 \, . 
\ee
However, $\Y^2<0$ for $n_2 \geq n_1$, thus, again, no solution occurs;

\item  $\X_1 = \X_2 = -\X_3$: Eqs.~\eqref{eq:anom3}--\eqref{eq:anom4} lead to
\be
\Y=\sqrt{\frac{n_2(n_2^2-1)+n_1(n_1^2-1)-2m(m^2-1)}{24m}} \, ,
\quad
\X_3 = \frac{3\alpha_{B+L}}{4m\Y^2}>0 \, . \label{eq:y1DL2MLcase1}
\ee
Solutions with (semi-)integer or rational values of the hypercharge $\Y$ can be systematically searched for by solving a generalized Pell's equation \cite{barbeau2003pell,scacco026diophantine}. 
An infinite subset of solutions with integer or semi-integer hypercharge values is found e.g.~with $n_2=m+k$ and $n_1=m-k$ (for $0<k\leq m-2$), i.e.
\be \label{eq:sol1DL2MLcase1}
\begin{split}
\Psi_1\sim\left(m-k, 0,-\frac{3\alpha_{B+L}}{k^2m}\right) \ ,
&\quad
\Psi_2\sim\left(m+k, 0,-\frac{3\alpha_{B+L}}{k^2m}\right) \ , \\
\Psi_3\sim\left(m, \frac{k}{2},\frac{3\alpha_{B+L}}{k^2m}\right) \ ,
&\quad
\Psi_4\sim\left(m, -\frac{k}{2},\frac{3\alpha_{B+L}}{k^2m}\right) \ .
\end{split}
\ee
If $m+k$ is odd (even), all the electric charges are integer (semi-integers).
Among this subset of solutions, the phenomelogical requirement (ii) holds only for $k=1$ and we checked by an explicit scan that there are no other solutions satisfying (ii) in the range $m,n_1,n_2\leq50$.

\item  $\X_1 = -\X_2 = \X_3$: Eqs.~\eqref{eq:anom3}--\eqref{eq:anom4} lead to
\be
\Y=\sqrt{\frac{n_2(n_2^2-1)-n_1(n_1^2-1)-2m(m^2-1)}{24m}} \, ,
\quad
\X_3 = \frac{3\alpha_{B+L}}{4m\Y^2}>0 \, . \label{eq:y1DL2MLcase2}
\ee
Again, solutions with (semi-)integer or rational values of the hypercharge $\Y$ can be systematically searched for by solving a generalized Pell's equation \cite{barbeau2003pell,scacco026diophantine}. 
An infinite subset of solutions with integer or semi-integer hypercharge values is found e.g.~with ansatz
$n_2=n_1+2m$, i.e.
\be \label{eq:sol1DL2MLcase2}
\begin{split}
\Psi_1\sim\left(n_1, 0,\frac{3\alpha_{B+L}}{m(m+n_1)^2}\right) \ ,
&\quad
\Psi_2\sim\left(n_1+2m, 0,-\frac{3\alpha_{B+L}}{m(m+n_1)^2}\right) \ , \\
\Psi_3\sim\left(m, \frac{m+n_1}{2},\frac{3\alpha_{B+L}}{m(m+n_1)^2}\right) \ ,
&\quad
\Psi_4\sim\left(m, -\frac{m+n_1}{2},\frac{3\alpha_{B+L}}{m(m+n_1)^2}\right) \ .
\end{split}
\ee
However, we checked via an explicit scan that no solutions satisfying (ii) are present in the range $m,n_1\leq50$.
\end{itemize}

\subsection*{Generalization to colored anomalon fields}
\label{app:gencolanom}

Finally, we briefly generalize the discussion to anomalon fields charged under $\SU(3)_C$. In this case, we only consider ML fields with real representation under the color gauge group and DL pairs whose color representations are each other's conjugate.

We note that the EW mixed anomaly cancellation conditions, which now read as
\begin{align}
\label{eq:anomcolor3}
[\SU(2)^2_L \U(1)_X] &:\quad \sum_\alpha \frac{n_\alpha(n_\alpha^2 -1)}{12} N_\alpha \X_{\alpha} + \frac{3}{2}\alpha_{B+L} = 0 \, , \\ 
\label{eq:anomcolor4}
[\U(1)^2_Y \U(1)_X] &:\quad \sum_\alpha n_\alpha N_\alpha \Y_{\alpha}^2 \X_{\alpha}   
- \frac{3}{2}\alpha_{B+L} = 0 \, , \\
\label{eq:anomcolor5}
[\U(1)_Y \U(1)^2_X] &:\quad  \sum_\alpha n_\alpha N_\alpha \Y_{\alpha}  \X_{\alpha}^2
= 0 \, ,
\end{align}
with $N_\alpha$ the color multiplicity of the anomalon $\Psi_\alpha$, admit the same solutions of the colorless case once their $\U(1)_X$ charges are rescaled as $\X_\alpha \to N_\alpha \X_\alpha$.

However, the $\U(1)_X$ charges are constrained by an additional anomaly cancellation condition which reads
\be \label{eq:anomcolor6}
[\SU(3)^2_c \U(1)_X] : \sum_\alpha n_\alpha T(N_\alpha) \X_\alpha = 0 \, ,
\ee
where $T(N_\alpha)$ is the Dynkin index of the color representation $N_\alpha$. Hence, among all rescaled solutions we have found, we have to select those that satisfy Eq.~\eqref{eq:anomcolor6}. 

Labelling quantum numbers as $\SU(3)_C\times \SU(2)_L \times \U(1)_Y \times \U(1)_X$, for the 2DL case the simplest solution is
\be
\label{eq:sol2DLcolor}
\begin{split}
\Psi_1\sim\left(6,6,\frac{5}{2},\frac{10\X-\alpha_{B+L}}{60}\right) \, ,
&\quad
\Psi_2\sim\left(6,6,\frac{5}{2},\frac{-10\X-\alpha_{B+L}}{60}\right) \, , \\
\Psi_3\sim\left(8,5, 3,\frac{10\X+\alpha_{B+L}}{80}\right) \, ,
&\quad
\Psi_4\sim\left(8,5, 3,\frac{10\X-\alpha_{B+L}}{80}\right) \, ,
\end{split}
\ee
while for the 1DL--1ML case, the simplest solution is
\be\label{eq:sol1DL1MLcolor}
\Psi_1\sim\left(8,5,0,-\frac{\alpha_{B+L}}{48}\right) \, ,
\quad
\Psi_{2,3}\sim\left(10,2, \pm\frac{3}{2},\frac{\alpha_{B+L}}{60}\right) \, .
\ee
For the 1DL--2ML case, we found no viable colorless solution; consequently, allowing for colored anomalons does not yield any viable realization either.

Eventually, none of the explicit models above with colored anomalons appears compatible with both direct-search bounds and finite-naturalness considerations, since direct searches for colored states require an anomalon mass scale $M_\Psi \gtrsim 2~\text{TeV}$~\cite{ATLAS:2020syg}, while the naturalness criterion (further strengthened once color multiplicities are included) points to $M_\Psi \lesssim 1~\text{TeV}$.

\begin{small}

\bibliographystyle{utphys}
\bibliography{bibliography.bib}

\end{small}

\end{document}